\documentclass[12pt]{article}
\usepackage[T2A]{fontenc}
\usepackage[cp1251]{inputenc}
\usepackage[russian]{babel}

\def\hybrid{\topmargin 0pt      \oddsidemargin 0pt
        \headheight 0pt \headsep 0pt

       \textwidth 6.5in        
       \textheight 9in         
        \marginparwidth 0.0in
        \parskip 5pt plus 1pt   \jot = 1.5ex}
\catcode`\@=11
\def\marginnote#1{}

\newcount\hour
\newcount\minute
\newtoks\amorpm
\hour=\time\divide\hour by60 \minute=\time{\multiply\hour by60
\global\advance\minute by-\hour}
\edef\standardtime{{\ifnum\hour<12 \global\amorpm={am}%
        \else\global\amorpm={pm}\advance\hour by-12 \fi
        \ifnum\hour=0 \hour=12 \fi
        \number\hour:\ifnum\minute<10 0\fi\number\minute\the\amorpm}}
\edef\militarytime{\number\hour:\ifnum\minute<10
0\fi\number\minute}

\def\draftlabel#1{{\@bsphack\if@filesw {\let\thepage\relax
   \xdef\@gtempa{\write\@auxout{\string
      \newlabel{#1}{{\@currentlabel}{\thepage}}}}}\@gtempa
   \if@nobreak \ifvmode\nobreak\fi\fi\fi\@esphack}
        \gdef\@eqnlabel{#1}}
\def\@eqnlabel{}
\def\@vacuum{}

\def\draftmarginnote#1{\marginpar{\raggedright\scriptsize\tt#1}}

\def\draftlabel#1{{\@bsphack\if@filesw {\let\thepage\relax
   \xdef\@gtempa{\write\@auxout{\string
      \newlabel{#1}{{\@currentlabel}{\thepage}}}}}\@gtempa
   \if@nobreak \ifvmode\nobreak\fi\fi\fi\@esphack}
        \gdef\@eqnlabel{#1}}
\def\@eqnlabel{}
\def\@vacuum{}
\def\draftmarginnote#1{\marginpar{\raggedright\scriptsize\tt#1}}

\def\draft{\oddsidemargin -.5truein
        \def\@oddfoot{\sl preliminary draft \hfil
        \rm\thepage\hfil\sl\today\quad\militarytime}
        \let\@evenfoot\@oddfoot \overfullrule 3pt
        \let\label=\draftlabel
        \let\marginnote=\draftmarginnote
   \def\@eqnnum{(\theequation)\rlap{\kern\marginparsep\tt\@eqnlabel}%
\global\let\@eqnlabel\@vacuum}  }

\def\numberbysection{\@addtoreset{equation}{section}
        \def\theequation{\thesection.\arabic{equation}}}

\def\underline#1{\relax\ifmmode\@@underline#1\else
        $\@@underline{\hbox{#1}}$\relax\fi}

\def\titlepage{\@restonecolfalse\if@twocolumn\@restonecoltrue\onecolumn
     \else \newpage \fi \thispagestyle{empty}\c@page\z@
        \def\thefootnote{\fnsymbol{footnote}} }

\def\endtitlepage{\if@restonecol\twocolumn \else  \fi
        \def\thefootnote{\arabic{footnote}}
        \setcounter{footnote}{0}}  

\relax

\draft

\numberbysection \hybrid

\def\beq{\begin{equation}}
\def\eeq{\end{equation}}
\def\p{\partial}
\def\G{\Gamma}
\def\g{\gamma}
\def\a{\alpha}
\def\s{\sigma}

\def\F{\Phi}
\def\m{{\rm mod}\ }
\def\n{\bar n}
\def\A{{\cal A}}
\def\L{{\cal L}}
\def\bea{\begin{eqnarray}}
\def\eea{\end{eqnarray}}
\def\X{{\cal X}}
\newtheorem{theo}{Theorem}[section]

\newtheorem{cor}{Corollary}[section]
\newtheorem{lem}{Lemma}[section]

\begin{document}

\begin{titlepage}

\title{Two-dimensional Toda lattice, commuting difference operators
and holomorphic vector bundles.}

\author{I.Krichever \thanks{Columbia University,
Landau Institute for Theoretical Physics and ITEP, Moscow, Russia.
Research is supported in part by National Science Foundation under
the grant DMS-01-04621 } \and S.Novikov \thanks{University of
Maryland, College Park and Landau Institute for Theoretical
Physics, Moscow . Research is supported in part by National
Science Foundation under the grant DMS-00-72700}}
\date{}
\maketitle
\begin{abstract}
High rank solutions to the $2D$ Toda Lattice System are
constructed simultaneously with the effective calculation of
coefficients of the higher rank commuting ordinary difference
operators. Our technic is based on the study of discrete dynamics
of Tyurin parameters characterizing the stable holomorphic vector
bundles over the algebraic curves (Riemann Surfaces).
\end{abstract}
\end{titlepage}

\section{Introduction.}

The aim of this work is to consider
a circle of problems closely related to
the construction of algebro-geometric high rank solutions of the
two-dimensional ($2D$) Toda lattice
\beq\label{toda} \p^2_{\xi
\eta}\varphi_{n}=e^{\varphi_n-\varphi_{n-1}}-e^{\varphi_{n+1}-\varphi_{n}}.
\eeq
High rank solutions of the Kadomtsev-Petviashvili (KP)
equation
\beq\label{kp} 3u_{yy}=\left(4u_t-6uu_x+u_{xxx}\right)_x
\eeq
were constructed in the earlier work of the authors \cite{kn1}.
Their construction, as well as the notion of {\it  rank} of the
solutions was based on the theory of commuting ordinary differential
operators \cite{kr2}. In modern mathematical physics this theory
arisen as an algebraic by-product of the integration theory of
soliton  equations and the spectral theory of periodic finite-gap
linear differential operators \cite{dmn,kr1,dkn,sol,nov} initiated
in \cite{nov}.

The KP equation, as well as any other soliton equation, is a part of
the corresponding hierarchy of commuting nonlinear partial
differential equations. A solution $u=u(t)$ of the KP hierarchy is a
function  of an infinite set of variables
$t=(t_1=x,t_2=y,t_3=t,t_4,\ldots)$. Beginning with the work
\cite{nov}, the algebro-geometric solutions of integrable
(1+1)-systems of the KdV type are singled out by the constraint
that they are stationary solutions for one of the flows of
corresponding hierarchy. The algebro-geometric solutions of the KP
equation were found in the work \cite{kr1}. They are singled out
by the constraint that they are stationary for {\it two} flows of
the hierarchy $\p_{t_n}u=\p_{t_m}u=0$. This constraint is
equivalent to the existence of a pair of commuting ordinary
differential operators
\beq\label{para} [L_n,L_m]=0, \ \
L_n=\sum_{i=0}^n u_i\p_x^i,\ \ L_m=\sum_{j=0}^m v_j\p_x^j\ ,
\eeq
that commute also with the auxiliary Lax operators $\p_y-L_2$,
$\p_t-L_3$ for the KP equation found in \cite{zakh,dr},
\beq\label{laxkp}
L_2=\p_x^2+u, \ L_3=\p_x^3+{3\over 2}\ u\p_x+w
\, .
\eeq
Therefore, pairs of commuting operators define invariant
subspaces for the KP hierarchy. By definition, rank $r$ of a pair
of commuting operators is the number of linear independent common
eigenfunctions, i.e. the number of solutions of the equations
\beq\label{sov}
L_n\psi^i=E\psi^i,\ \ \ L_m\psi^i=w\,\psi^i,\ \
i=1,\ldots,r\ ,
\eeq
for $(E,w)$ such that a solution exists.
The rank of the pair of commuting differential operators is a
common divisor of orders of the operators.

The classification problem of commuting ordinary differential
operators with scalar coeffi\-cients was considered as a pure
algebraic problem by Burchnal and Chaundy in the 20-s
\cite{ch1,ch2}. They solved the problem for the case of operators
of co-prime orders (where the rank always equals 1) although no
explicit expressions for the coefficients of the operators were
found. Effective classification of commuting rank 1 operators were
obtained in \cite{kr1}. Burchnal and Chaundy emphasized (see
\cite{ch2}), that the problem for the case of rank $r>1$ is very
complicated.

The first steps were made in \cite{dix,drin}. A method of effective classification
of commuting generic differential operators of rank $r>1$ was developed by
the authors in \cite{kn1,kr2}. Commuting pairs of differential operators
rank $r$
depend on $(r-1)$ {\it arbitrary functions} of one variable, smooth algebraic curve
$\G$ with {\it one} puncture $P$ and a set of the Tyurin parameters
for a framed stable holomorphic vector bundle over $\G$. The corresponding
constructions are called {\it one-point} constructions.

For the $2D$ Toda lattice equations the constraints that single out
algebro-geometric solutions can be described in the same way as in
the KP theory. The condition that these solutions are stationary
for two flows of the hierarchy is equivalent to the existence of a
pair of commuting $\xi,\eta$-dependent {\it difference} operators
\beq \label{01}
L=\sum_{i=-N_-}^{N_+} u_i(n)T^i, \ \
A=\sum_{i=-M_-}^{M_+}v_i(n)T^i,
\eeq
that commute also with the
Lax operators
\beq\label{02}
\L_1=\p_{\xi}-T-w(n),\ \
\L_2=\p_{\eta}-c(n)T^{-1}, \eeq \beq\label{02a}
w(n)=\varphi_{n\xi},\ c(n)=e^{\varphi_n-\varphi_{n+1}},
\eeq for
the $2D$ Toda equations. Here and below $T$ denotes the shift
operator, $Ty_n=y_{n+1}$.

The well-known theory of commuting rank $r=1$ scalar difference
operators was made of only {\it two-point constructions}
\cite{kr3,mam}. A ring of such operators is isomorphic to the ring
$A(\G,P^{\pm})$ of meromorphic functions on an algebraic curve
$\G$ with the poles at two punctures $P^{\pm}$. Remarkable, {\it almost
graded} structure of these rings, their connections with the
Fourier-Laurent type basis on Riemann surfaces, was developed by
the authors and used for the string theory \cite{kn2}.

The classification problem of commuting difference operators in
its full generality has not been solved yet. Analysis of the problem
presented in section 2 of this article has allowed  the authors to
obtain some new important results. These results were announced
in \cite{kn,kn2t}. It turns out that commuting rank $r=1$
difference operators can be obtained with the help of {\it
multi-point} constructions with any number of infinite points
including one point if rank is even (which is most astonishing).
Multi-point constructions are well known in the theory of commuting
differential operators with matrix coefficients \cite{kr1}. It
seems reasonable to consider commuting difference operators
constructed with the help of only {\it one- and two-point}
constructions, as a difference analog of commuting differential
operators with scalar coefficients. For the authors the supporting
evidence of this suggestion is that only in these two cases the
rings of commuting operators are invariant with respect to $2D$
Toda hierarchy.

It is necessary to emphasize that one- and two-point
constructions differ drastically. There are no functional
parameters in the two-point case. Coefficients of the
corresponding operators can be written in terms of the Riemann
theta-functions. The one-point construction of commuting operators
of rank $l>1$ contains {\it functional parameters}. The
coefficients of commuting operators depend on $l$ arbitrary
functions of discrete variable, smooth algebraic curve $\G$
with {\it puncture} $P$ and a set of Tyurin parameters. As in
continuous case  \cite{kr2}, the reconstruction problem of
the eigenfunctions of the corresponding operators is equivalent
to the Riemann problem on $\G$, and in general can not be solved
explicitly. At the same time, as it was found in \cite{kn1},
in some cases the
reconstruction problem for the coefficients of the commuting
operators  can be solved explicitly. The solution is
based on equations for the Tyurin parameters describing their
dependence on normalization point. Note, that in the recent work of
one of the authors \cite{kr-hit}, connections of continuous
deformations of the Tyurin parameters, describing high rank
solutions of the KP equation, and the Hitchin systems were
established \cite{hitchin}. Using a discrete dynamics of the
Tyurin parameters, the authors found explicit form of rank 2
commuting operators corresponding to an elliptic curve with {\it
one} puncture. The ring of such operators is generated by a pair
of operators $L$ and $A$ of orders $4$ and $6$, respectively:
$$
L=\sum_{i=-2}^2u_i(n)T^i, \ \ A=\sum_{i=-3}^3v_i(n)T^i\, .
$$
The coefficients of the operators depend on two arbitrary functions
$\g_n$ and $s_n$. In the simplest case an explicit form of $L$ is
given by the formulae:
$$ L=L_2^2-\wp(\g_n)-\wp(\g_{n-1}), $$
where $L_2$ is the difference Schr\"odinger operator
$$
L_2=T+v_n+c_nT^{-1} $$ with the coefficients
\bea
4c_{n+1}&=&(s_n^2-1)F(\g_{n+1},\g_n)F(\g_{n-1},\g_n),\nonumber\\
2v_{n+1}&=&s_n F(\g_{n+1},\g_n)-s_{n+1}
F(\g_{n},\g_{n+1})\nonumber. \eea where
$$
F(u,v)=\zeta(u+v)-\zeta(u-v)-2\zeta(v) .
$$
Here and below
$\wp(u)=\wp(u|\omega, \omega'),\ \zeta(u)=\zeta(u|\omega,\omega')$
are classical Weiershtrass functions.

These formulae are discrete analogs of the formulae for
differential operator $L_4$ of the order 4 and rank  2, obtained by the authors in
\cite{kn5}.
In \cite{gr1} an explicit expression of the functional parameter in the formulae
for $L_4$ corresponding to the Dixmier operator \cite{dix} was found.
Coefficients of this operator are polynomial functions of the independent variable
$x$. It would be interesting to find a discrete analog of the Dixmier
operator.

The construction of high rank algebro-geometric solutions of the $2D$ Toda lattice
was briefly described in \cite{kn,kn2t}. We discuss it in detail in Section 3.
In is based on the construction of the vector Baker-Akhiezer function as
a deformation of the eigenfunctions of commuting differential operators. These
deformations are defined by the form of the Baker-Akhiezer functions
in the neighborhood of punctures. The form is defined with the help of some
{\it germ} matrix functions. Once again it is necessary to stress the difference
between one and two puncture cases.

In the two-point case (in which there are no functional parameters in the
construction of commuting differential operators), the Baker-Akhiezer
function is defined by two auxiliary germ functions
$\Psi_{+}(\xi,z), \ \Psi_-(\eta,z)$. Each of them is defined by an ordinary
differential equation containing arbitrary functions of
one of the continuous variables $\xi$ or $\eta$, respectively.
In the one-puncture case, functional parameters are
present in the construction of the corresponding commuting operators, but
there is no ambiguity in the definition of the germ function $\Psi_0(n,\xi,\eta,z)$.

In both the cases the multi-parametric Baker-Akhiezer functions
$\psi_n=(\psi_n^i)$
satisfy the equations
\beq\label{au}
\L_1\psi_n=0, \ \ \L_2\psi_n=0,
\eeq
where $\L_1,\ \L_2$ are operators of the form (\ref{02}).
Compatibility of (\ref{au}) is equivalent to (\ref{toda}). Equations (\ref{au})
imply also that for each $n$ the Baker-Akhiezer function, as a function of the
variables $(\xi,\eta)$, satisfies the linear Schr\"odinger equation:
\beq\label{shred}
\left(\p^2_{\xi,\eta}+v_n(\xi,\eta)\p_{\xi}+u_n(\xi,\eta)\right)\psi_n=0.
\eeq
Therefore, our constructions provide at the same time the construction
of two-dimensional Schr\"odinger operators in the magnetic field that are
integrable on one energy level.

Inverse spectral problem on \ {\it one energy level} for the Schr\"odinger
operators in a magnetic field was introduced and solved
in  \cite{dkn1} for the rank 1 case. The construction of
$2D$ Schr\'odinger operators of arbitrary rank integrable on one energy
level  was proposed in
\cite{kn5}. All these constructions are {\it two-point} constructions.
The possibility of the one-point construction of integrable Schr\'odinger operators
has never been discussed until now .

We would like to stress that the problem of construction of
solutions of the $2D$ Toda lattice equations is equivalent to the
problem of construction of integrable Schr\"odinger operators in a
magnetic field. In the modern theory of integrable systems
equations (\ref{toda}) and its Lax representation were obtained in
the framework of the Zakharov-Shabat scheme as two dimensional
analog of the one-dimensional Toda lattice (\cite{Mi}). Later it
became clear that these equations had been known in the classical
differential geometry in an equivalent form of chains of the
Laplace transformations for the two-dimensional Schr\"odinger
operator (\cite{DT}).

Let us consider the two-dimensional Schr\"odinger operator $L$ in a
magnetic field
\beq\label{sh1} 2L=(\p_{\bar z} +B)({\p_z} + A) +
2V ,
\eeq
where $\p_z = \p_x - i\p_y$, $\p_{\bar z} = \p_x +
i\p_y$.
Usually the functions $H=(B_z - A_{\bar z})/2$ и $U=-H-V$
are called {\it magnetic field \/} and {\it potential\/},
respectively. Below, for brevity, the function $V$ would be called
the potential of $L$. The operator  $L$ is defined up to the gauge
transformation $L\rightarrow e^{-f}Le^{f}$. The only invariants of
$L$ are the potential $V$ and the magnetic field $H$. {\it Laplace
transformation} is a two-dimensional analog of the
B\"aclund-Darboux transformation. It is defined as follows
\beq\label{sh5} (\p_{\bar z} +B )(\p_z+ A)+2V \ \ \ \longmapsto\ \
\ \widetilde  V(\p_z+ A)V^{-1}(\bar\p_z+B)+2V. \eeq The potential
$\widetilde V$ and the magnetic field $\widetilde H$ of the
transformed operator are given by the formulae
\beq\label{sh6}
\widetilde V=V+\tilde H,\qquad 2\widetilde H = 2H+\Delta \ln V.
\eeq
If a function $\psi$ satisfies the equation $L\psi=0$, then
the function $\widetilde \psi=\left(\p_z+A\right)\psi$ is a
solution of the equation  $\widetilde L\ \widetilde \psi=0$. Let
us consider a chain the Laplace transformations \beq\label{sh8}
V_{n+1}=V_n+H_{n+1},\ \ 2H_{n+1}=2H_n+\Delta \ln V_n. \eeq After
the change of variables $V_k=\ln
\left(\varphi_n-\varphi_{k-1}\right)$ the chain of the Laplace
transformations becomes equivalent to the the $2D$ Toda lattice
equations (\ref{toda}).

\section{Commuting difference operators}

\subsection{Formulation of the problem.}
The main goal of this section is to construct an effective classification
of commuting
\beq\label{3}
[L,A]=0
\eeq
scalar difference operators of the form
\beq \label{1}
L=\sum_{i=-N_-}^{N_+} u_i(n)T^i, \ \ A=\sum_{i=-M_-}^{M_+}
v_i(n)T^i\, , \ \ N_{\pm}=r_{\pm}n_{\pm},\ \ M_{\pm}=r_{\pm}m_{\pm},
\eeq
where $r_{\pm}$ are maximum common divisors of
maximum and minimum orders of the operators, respectively, i.e. the non-zero numbers
$n_{\pm}, m_{\pm}$ are co-prime:
\beq\label{2a} (n_+,m_+)=(n_-,m_-)=1.
\eeq
Let us assume, that the leading coefficients of these operators are different
from zero. For brevity, they are denoted by
\beq\label{2}
u^{\pm}(n)=u_{\pm
N_{\pm}}(n)\neq 0,\ \ v^{\pm}(n)=v_{\pm M_{\pm}}(n)\neq 0.
\eeq
Equation (\ref{3}) is invariant under the gauge transformation
\beq\label{4}
L,A\ \longmapsto \tilde L=gLg^{-1}, \tilde A=gAg^{-1}, \ \
\tilde L=\sum_{i=-N_-}^{N_+} g(n+i)g^{-1}(n)u_i(n)T^i,
\eeq
where $g(n)\neq 0$ is an arbitrary non-vanishing function of the discrete variable
$n$. This transformation can be used for the following normalization of $L$
\beq \label{5}
u^{+}(n)=1.
\eeq
The normalization (\ref{5}) which will be always assumed below, is invariant under
the gauge transformations corresponding to the functions $g(n)$ such that
$g(n+N_+)=g(n)$. Equation (\ref{3}) implies that the leading coefficient of the
operator $A$ satisfies the equation $v^+(n+N_+)=v^+(n)$.
Therefore, the gauge transformations can be used in order to fix
the normalization
\beq \label{7}
v^+(n+r_+)=v^+(n)=v^+_{\n}
\eeq
in addition to (\ref{5}).
Here $\n$ is a residue of $n$ mod $r_+$, i.e.
$n\to \n=n(\m r_+),\ 0\leq \n <r_+-1.$
The normalization (\ref{5},\ref{7}) is invariant under
the gauge transformations corresponding to the functions  $g(n)$
such that $g(n+r_+)=g(n)$.

If the normalization (\ref{5},\ref{7}) is fixed, then for the
coefficients on the other edge have the form
\beq\label{7b} u^-(n)=h^{-1}(n-N_-)h(n), \ \
v^-(n)=v^-_{\tilde n}h^{-1}(n-M_-)h(n). \eeq
Here $\tilde n=n(\m r_-)$.

In the case $r_+=r_-=r>1$ the commutativity equations have additional symmetry.
Namely, let $y_n$ be a function of the discrete variable $n$, then for
any $0 \leq i<r$ we define a new function $Y_n$
\beq
Y_n =\left\{\begin{array}{ll} 0,&n\neq i\ (mod \ r)\\ y_{\bar n}, &
n=r\bar n+i\end{array}\right.
\eeq
Under that correspondence a difference operator of order $N$ defines an operator
of order $rN$. The direct sum of $r$ pairs of commuting operators
$[L_i,A_i]=0$ considered as operators acting on defined above sublattices,
satisfies a non-degeneracy conditions (\ref{2}) and the commutativity
equation (\ref{3}).

Below we will consider only commuting pairs of {\it irreducible} operators, i.e.
operators that have no block-diagonal or jordan-cell form under decomposition
of the $n$-lattice into a sum of sublattices $kr+i,\
0\leq i<r-1$. A pair of commuting operators is irreducible if, for example,
for each $i=1,\ldots, r-1$ there is
$n_0$ so that
\beq\label{nep} v_{rm_+-i}(n_0)\neq 0, \
v_{-rm_-+i}(n_0)\neq 0.
\eeq

\subsection{Spectral curve. Formal infinities.}
In the discrete case an {\it affine} spectral curve is defined identically
to the continuous case. Let us consider a linear space
${\cal L}(E)$ of solutions of the equation $Ly=Ey$, i.e.
\beq \label{eig0}
y_{N_+}+\sum_{i=-N_-}^{N_+-1} u_i(n)y_{n+i}=Ey_n.
\eeq
Dimension of this space is equal to the order of
$L$, $\dim {\cal L}(E)=N_++N_-$. Restriction of  $A$
onto ${\cal L}(E)$,
\beq
\label{8} A(E)=A|_{{\cal L}(E)},
\eeq
is a finite-dimensional linear operator. The spectral curve, which parameterizes
common eigenfunctions of $L$ and $A$, is defined by the characteristic equation
\beq \label{sp} R(w,E)=\det(w-A(E))=0.
\eeq
Let $c^i$ be a basis  of solutions of (\ref{eig0}),
defined by the initial conditions
\beq\label{nac}
c_n^i=\delta^i_n,\ \ i,n=-N_-,\ldots, N_+-1\, .
\eeq
Matrix entries of  $A(E)$ in that basis
are polynomial functions of the variable $E$. Therefore, $R(w,E)$
is a polynomial in both the variables $w$ and $E$.

\medskip
\noindent
{\bf Compactification of the spectral curve.}
In the discrete case a construction of common eigenfunctions of commuting
operators at infinity  $E\to \infty$ is on the whole parallel to the continuous
case, but at the same time contains some new important features.

Let $\L_+$ be the linear space of solutions of the equation
\beq \label{eig} L\F=z^{-N_+}\F, \ \ \ \F=\{\F_n\}
\eeq
of the form
\beq
\label{Psi} \F_n(z)=z^{-n}\left(\sum_{s=0}^{\infty} \xi_s(n)z^s\right).
\eeq
It is assumed that $\xi_0(n)$ is not equal to zero for at least one value of $n$.
\begin{lem} The space $\L_+$, considered as a linear space over the field $k_+$ of
Laurent series in the variable $z$, is of dimension $N_+$. It is spanned
by the solutions $\F^i, \ i=0,\ldots, N_+-1$, uniquely defined by the conditions
\beq \label{Psi1}
\F_n^i(z)=z^{-n}\delta_{ni}, \ \ n,i=0,\ldots, N_+-1. \eeq
\end{lem}
For the proof of this Lemma it is enough to note that a substitution
of the series (\ref{Psi}) into (\ref{eig}) gives a recurrent system of equations for
the coefficients of (\ref{Psi})
\begin{eqnarray}
\xi_0(n+N_+)&=&\xi_0(n), \nonumber \\
\xi_1(n+N_+)&=&\xi_1(n)-u_{n+N_+-1}(n)\xi_0(n+N_+-1),\ \ ...
\end{eqnarray}
The system implies that $\F$ is uniquely defined by the initial data $\xi_s(i), \ i=0,\ldots, N_+-1$.

The space $\L_+$ is invariant with respect to the operator $A$. Therefore,
\beq\label{a}
A\F_n^i(z)=\sum_{j=0}^{N_+-1}A_{ij}(z)\F^j_n(z), \ \
A_{ij}(z)=\sum_{s=-M_+}^{\infty} A^{(s)}_{ij}z^s. \eeq
The leading coefficient is a matrix
\beq\label{st}
A^{(M_+)}_{ij}=v^+_i\delta_{j,\,i+M_+(\, mod \ N_+)} \eeq
The change $z\to \varepsilon z$, where $\varepsilon^{N_+}=1$, defines an automorphism
of $\L_+(z)$. That implies that coefficients of the characteristic polynomial
\beq\label{sp2}
\det(w-A_{ij}(z))=w^{N_+}+\sum_{i=0}^{N_+-1} a_i(E)w^i,\
E=z^{-N_+},
\eeq
are series in the variable $E=z^{-N_+}$. From that it follows that the eigenvalues
are Laurent polynomials in the variable  $z$
(but not in fractional powers $z^{1/k}, k>1$).

Let us assume that
\beq\label{r1} v^{+}_i\neq v^{+}_j, \ i\neq j , \eeq
where $v^+_i$ are defined in (\ref{7}). In that case the matrix $A_{ij}(z)$ has
a unique eigenvector $\Psi_n^{(i)}(z),\ i=0,\ldots,r_+-1,$
of the form (\ref{Psi})
\beq \label{Psi1a}
\Psi^{(i)}_n(z)=z^{-n}\left(\delta_{i,\n}+\sum_{s=1}^{\infty}
\xi_s(n)z^s\right). \eeq
The leading term of the expansion is an eigenvector of the matrix (\ref{st}),
corresponding to its eigenvalue $v^+_i$. Under the change $z\to \varepsilon^{n_+} z$,
the leading term gets the factor $\varepsilon^{-in_+}$. Therefore, the vector
$z^i\Psi^{(i)}_n(z)$ has an expansion in powers
\beq \label{z1}
z_0=z^{r_+}=E^{1/n_+}.
\eeq
That implies for  $i>0$ and $j=0,\ldots, r_+-1,$ the following equalities:
\bea \label{Psi2}
\Psi^{(i)}_{kr_++j}(z)&=&z^{-i}z_0^{-k}(O(z_0)), \ \ j<i \, ,\nonumber\\
\Psi^{(i)}_{kr_++j}(z)&=&z^{-i}z_0^{-k}(O(1)), \ \ j\geq i. \eea
From the irreducibility condition (\ref{nep}), in which without loss of generality
we put $n_0=0$, it follows that for $i>0$ the zero component of the eigenvector
has the form $\Psi_0^{(i)}=O(z^{r-i})$. Therefore, the normalized
eigenvectors
\beq \label{Psi3}
\psi_n^{(i)}={\Psi_n^{(i)}(z)\over \Psi_0^{(i)}(z)}, \ \
\psi_0^{(i)}=1, \eeq
for $i>0$ have the form
\begin{eqnarray} \label{psi}
\psi^{(i)}_{kr_++j}(z_0)&=&O(z_0^{-k-1}), \ \ i\leq j\, , \nonumber \\
\psi^{(i)}_{kr_++j}(z_0)&=&O(z_0^{-k}),  \ \  j<i.
\end{eqnarray}
For $i=0$ we have:
\beq \label{psi0}
\psi^{(0)}_{kr_++j}(z_0)=z_0^{-k}(1+O(z_0)).
\eeq
Hence, in the case when the conditions (\ref{r1}) are satisfied, we have constructed
a set of  $N_+$ formal eigenvectors of the operator $A(E)$ in the neighborhood
of the infinity $E=\infty$. Below, we construct
an additional set of $N_-$ formal eigenfunctions under the assumption
\beq\label{r2}
v^{-}_i\neq v^{-}_j, \ i\neq j , \eeq
where $v^-_i$ are defined in (\ref{7b}).

Let us consider a linear space $\L_-$ over the field of Laurent series $k_-$
in the variable $z_-$, spanned by the solutions of the equation
\beq \label{eig-}
L\F=z^{-N_-}\, \F \eeq
of the form
\beq \label{Psi-}
\F^-_n(z)=z_-^{n}\left(\sum_{s=0}^{\infty} \xi^-_s(n)z_-^s\right).
\eeq
It is assumed that $\xi^-_0(n)$ is not equal to zero for at least one $n$.

As before, we obtain that the characteristic polynomial of the operator
$A^-(z_-)$, induced by an action of $A$ on $\L_-$, has the form
\beq
\label{char-} \det(w-A^-(z_-))=
\prod_{i=0}^{r_-}\prod_{k=0}^{n_-}\left(w-\hat v^-_i(z_{k,-})\right),
\ \ z_{k,-}=\varepsilon_1^k z_-^r, \eeq
where
$\varepsilon_{1}^{n_-}=1$, and the series $\hat v^-_i=\hat v^-_i(z_{0,-})$
has the form:
\beq \label{v} \hat v^-_i(z_{0,-})=v^-_i
z_{0,-}^{-m_-}(1+O(z_{0,-}))\ .
\eeq
The normalized eigenvector of $A^-(z_-)$, corresponding to the eigenvalue
$w=\hat v^-_i(z_{0,-})$ for all $i=0,\ldots,r-1,$ has the form:
\begin{eqnarray} \label{psi-}
\psi^{(i)}_{kr_-+j}(z_{0,-})&=&O(z_{0,-}^{k}), \ \ i\geq j\, , \nonumber \\
\psi^{(i)}_{kr_-+j}(z_{0,-})&=&O(z_{0,-}^{k+1}),  \ \  i<j.
\end{eqnarray}
In the direct sum of the spaces $\L_+$ and $\L_-$ it is possible to choose a basis
$c^{i}_n$, normalized by the conditions
\beq
c^{i}_n=\delta_{i,\, n}, \ \ i,n=N_-,\ldots,N_+-1\, . \eeq
In this basis the operator $A$ has the same matrix entries as in the construction
above of the affine spectral curve. Therefore, the characteristic equation (\ref{sp})
coincides with the product of the characteristic equations for $A(z)$ and $A^-(z_-)$,
i.e.
\beq
\label{char10} \det (w-A(E))=
\left[\prod_{i=0}^{r_+}
\prod_{k=0}^{n_+-1}\left(w-\hat v_{i}(z_{k})\right)\right]
\left[\prod_{i=0}^{r_-}\prod_{k=0}^{n_--1}\left(w-\hat
v^-_{i}(z_{k,-})\right)\right]\, ,
\eeq
where the products in the first and the second sets of factors are taken over
all roots of $E^{-1}$ of degree $n_+$ and $n_-$, respectively:
\beq
E=z_k^{-n_+}=z_{k,-}^{-n_-}
\eeq
The expansion (\ref{char10}) gives a complete information on compactification of the
spectral curve when the operator $A(E)$ has $N$ distinct eigenvalues
for almost all $E$. It is the case of commuting operators of rank 1, which
is considered in the next section.

\subsection{Commuting operators of rank 1.}

The conditions (\ref{r1}) are (\ref{r2}) imply that the eigenvalues of $A(E)$,
corresponding to factors of the first and in the second products in the formula
(\ref{char10}) are distinct. In addition, let us require that the factors in
the two products do not coincide with each other. For that it is sufficient to assume
one of the following conditions
\beq\label{r3}
(i)\ \ m_+n_-\neq m_-n_+ ,\ \ \
(ii)\ \ v^+_i\neq v^-_j\, .
\eeq
In this case equation (\ref{char10}) implies that the affine spectral curve
defined by equation (\ref{sp}), is compactified by
$l=(r_++r_-)$ points $P_{i_{\pm}}^{\pm}, \ \ i_{\pm}=0,\ldots, \, r_{\pm}-1. $
In the neighborhood of the infinity, and therefore for almost all
$E$ the spectral curve $\G$ has $N=N_++N_-$ sheets. Therefore, to every point
of the spectral curve there correspond the unique up to multiplication
common eigenfunction $\psi_n$ of the operators $L$ and $A$.
Let us present the main theorem.

\begin{theo} Let irreducible commuting operators satisfy the conditions
(\ref{r1}, \ref{r2}, \ref{r3}). Then:

(1) the spectral curve $\G$, given by the characteristic equation (\ref{sp})
is compactified at the infinity by $l=(r_++r_-)$ points $P_{i_{\pm}}^{\pm}$,
in the neighborhoods of which local coordinates are:
$$ z_{k,\pm}=E^{-1/n_{\pm}};$$

(2) the common eigenfunction of the commuting operators
$$L\psi_n(Q)=E\psi_n(Q),\ \ A\psi_n(Q)=w\psi_n(Q), \ \ Q=(w,E)\in \G,$$
normalized by the condition $\psi_0=1$ is a meromorphic function on $\G$,
with the divisor of poles $D=\{\g_s\}$ outside the punctures $P_{i}^{\pm}$
does not depend on $n$. In the neighborhoods of the punctures
$\psi_n$ has the form (\ref{psi},\ref{psi0},\ref{psi-}), i.e. if \  $n$ is represented
in the form $n=kr_++j_+=k'r_-+j_- ,\, 0\leq j_{\pm}<r_{\pm}$, then

(2a) $\psi_{n}$ has the poles of order $k$ at the punctures
$P_0^+,P^+_{j_++1},\ldots, P_{r_+-1}^+$ and the poles of order $k+1$  at
the punctures $P_1^+,\ldots, P_{j_+}^+$;

(2b) $\psi_{n}$ has zeros of order $k'$ at the punctures
$P_{j_-}^-,\ldots,P_{r_--1}^-$ and zeros of order $k'+1$  at the punctures
$P_0^-,\ldots,P_{j_--1}^-$

(3) in the general position, when the spectral curve is smooth and irreducible,
the number of poles $\g_s$ (counting with their multiplicities) of
$\psi_n(Q)$ outside the punctures is equal to the genus $g$ of the curve $\G$.
\end{theo}
The theorem defines a map which for a generic pair of commuting operators,
satisfying the conditions of the theorem, assigns a smooth algebraic curve $\G$ with $l$
punctures $P_{i_{\pm}}^{\pm}$ and a divisor of degree $g$:
\beq
[L,A]=0 \longmapsto \{\G, P_{i_{\pm}}^{\pm}, D=\{\g_s\}\}, \
0\leq i_{\pm}<r_{\pm},\ s=1,\ldots, g. \label{ag}
\eeq
Let us show that these algebro-geometric data define uniquely the commuting
operators.

Let $\G$ be a smooth genus $g$ algebraic curve with $l=(r_++r_-)$ punctures
$P_{i_{\pm}}^{\pm}$. From the Riemann-Roch theorem it follows that for
any non-special divisor $D=(\g_1,\ldots,\g_g)$ there is a
function $\psi_{n}(Q)$ unique up to a factor,
such that its divisor of the poles outside the punctures
is not greater than $D$, and such that at the punctures $P_i^{\pm}$ it has poles
and zeros of the orders defined in $(2a, 2b)$.

Indeed, the conditions $(2a, 2b)$ mean that $\psi_{n}(Q)$ belongs to a space
of meromorphic functions $\L(D_{n})$, associated with the divisor $D_{n}$
$$D_{n}=D+k\sum_{i_+=0}^{r_+-1} P_{i_+}^++\sum_{i=1}^{j_+}P_{i_+}^+
-k'\sum_{i_-=0}^{r_--1} P_{i_-}^-
-\sum_{i_-=0}^{j_--1}P_{i-}^-\, ,$$
where $n=kr_++j_+=k'r_-+j_-$.
This divisor is of degree $g$, therefore, according to the Riemann-Roch theorem
the space  $\L(D_{n})$ is one-dimensional.

Let us denote a ring of the meromorphic functions on
$\G$ with poles at $P_{i_{\pm}}^{\pm}$ by $\A(\G, P_{i_{\pm}}^{\pm})$.

\begin{theo} Let $\psi(Q)=\{\psi_n(Q)\}$ be a sequence of functions, corresponding
to algebro-geometric data (\ref{ag}). Then, for each function
$f\in \A(\G, P_i^{\pm})$ there is a unique difference operator $L_f$
(with coefficients independent of $Q$), such that
$$ L_f\psi(Q)=f(Q)\psi(Q).$$
If the function $f(Q)$ has poles of order $n_+$ and $n_-$ at the punctures
$P_{i_{\pm}}^{\pm}$ respectively, then the operator $L_f$ has the form (\ref{1}).
\end{theo}
The proof follows immediately from the Riemann-Roch theorem.
If the function $f$ has poles at $P_i^{\pm}$ of order $n_{\pm}$, then the
function $f(Q)\psi_n(Q)$  belongs to the linear space
$$f(Q)\psi_n(Q)\in \L
\left(D_n+n_+\sum_{i_+=0}^{r_+-1}P_{i_+}^++n_-\sum_{i_-=0}^{r_--1}P_{i_-}^-\right).$$
This space is of the dimension $N_++N_-+1$. From the definition of $\psi_n$
it follows, that the functions $\psi_{n+i}, \ -rn_-\leq i\leq rn_+,$ are the basis
of this space. Coefficients $u_i(n)$ of the operator $L_f$ are just coefficients of expansion of
$f\psi_n$ in the basis of the functions $\psi_{n+i}$.

The function $\psi_n(Q)$ can be written explicitly in terms
of the Riemann theta-functions. For simplicity we present the corresponding
formulae for the case $r=r_+=r_-$.
Let $a_i^0,b_i^0$ be a basis of cycles on $\Gamma$ with canonical matrix of
intersections $a_i^0\cdot a_j^0=b_i^0\cdot b_j^0=0, \ a_i^0\cdot
b_j^0=\delta_{ij}$. In a standard way it defines the basis of normalized
holomorphic differentials $\omega_j(P)$.
The matrix $B$ of their $b$-periods defines the Jacobian variety $J(\Gamma)$ and
the Riemann theta-function $\theta(z)=\theta(z|B)$.
Let us introduce the functions
$h_j,\  j=0,\ldots, r-1$,
\beq
h_j(Q)={f_{j}(Q)\over f_{j}(P_j^+)}; \quad
f_{j}(Q)={\theta(A(Q)+Z_{j})\over \theta (A(Q)+Z_0)}
{\prod_{i=0}^{j-1}\theta(A(Q)+S^-_i)\over \prod_{i=1}^j\theta
(A(Q)+S_i^+)}, \label{2.260}
\eeq
where
\beq
S^{\pm}_i={\cal K}-A(P_i^{\pm})-\sum_{s=1}^{g-1} A(\gamma_s), \ \ i=0,\ldots,r-1.
\eeq
\beq
Z_{j}=Z_0+\sum_{i=0}^{j-1}A(P_{i}^-)-\sum_{i=1}^{j}A(P_{i}^+)\quad
Z_0={\cal K}-\sum_{s=1}^{g} A(\gamma_s),
\label{2.263}
\eeq
and $A(Q)$ is a vector with the coordinates $A_k(Q)=\int_{q_0}^Q \omega_k$.
Let $d\Omega^{(0)}$ be a unique meromorphic differential on $\G$, with
simple poles at $P_j^{\pm}$ with the residues $\mp 1$,
and normalized by the conditions
\beq
\oint_{a_k^0}d\Omega^{(0)}=0. \label{2.24}
\eeq
The coordinates of the vector of its $b_0$-periods $U^{(0)}$ equal
\beq U^{(0)}_k={1\over 2\pi i} \oint_{b_k^0} d\Omega^{(0)}=
\sum_{j=0}^{r-1}\left(A(P_j^{-})-A(P_j^+)\right) \label{2.25}
\eeq
\begin{lem}
The function $\psi_n(Q)$ equals
\beq
\psi_{rk+j}=h_{j}(Q) {\theta
\left(A(Q)+kU^{(0)}+Z_{j}\right) \theta \left(A(P_j^+)+Z_{j}\right)
\over \theta \left(A(Q)+Z_{j}\right)
\theta \left(A(P_j^+)+kU^{(0)}+Z_{j}\right) } \exp{\left(k\int^Qd\Omega^{0}\right)}
\label{2.29}
\eeq
\end{lem}
For the proof of (\ref{2.29}) it is enough to
show that a function defined by the formula is single-valued on $\G$ and has all
the required analytical properties.
\begin{cor} The coefficients of commuting rank 1 generic
operators are quasi-periodic meromorphic functions of the variable $n$.
\end{cor}

\subsection{Rank $> 1$. The case of separated infinities.}
From the construction of formal common eigenfunctions of the operators $L$ and $A$
it follows, that for almost all values of $E$ the operator $A(E)$ is diagonalizable.
We call a set of the multiplicities $\mu=\left(\mu_1,\ldots,\mu_k\right)$ of
the eigenvalues of $A(E)$ vector rank of the commuting operators.
Note, that rank of commuting differential operators with scalar
coefficients is always scalar $(k=1)$, and is a common divisor of
orders of the operators \cite{kr2}. The notion of vector rank
in the classification problem of commuting differential operators with matrix
coefficients was introduced in в \cite{gr}.
For commuting operators of a vector rank $\mu$ the characteristic equation has
the form
\beq\label{rang}
\det(w-A(E))=\prod_{i=1}^k R_i^{\mu_i}(w,E).
\eeq
We would like to emphasize that conditions (\ref{r1}, \ref{r2}), which
are sufficient for simplicity of the eigenvalues of the operators $A(z), A^-(z_-)$
in the formal neighborhood of the infinity are incompatible with
auxiliary linear problems for the $2D$ Toda lattice equations.
Indeed, from the equations
\beq\label{rang1}
[L,\L_i]=[A,\L_i]=0,
\eeq
where $L,A$ и $\L_i$ have the form (\ref{01}) и (\ref{02}) it follows that
\begin{eqnarray}
u_{N_+}(n+1)=u_{N_+}(n),\ u_{N_-}(n)c(n-N_-)=u_{N_+}(n-1)c(n),\nonumber\\
v_{M_+}(n+1)=v_{M_+}(n), \
v_{M_-}(n)c(n-N_-)=v_{M_+}(n-1)c(n). \nonumber
\end{eqnarray}
The later equations imply that  $v_i^{\pm}$, given by
(\ref{7}, \ref{7b}), satisfy the equations
\beq\label{rang3}
v_i^+=v^+, \ \ v_i^-=v^-,
\eeq
In that case, {\it a priori} there are no obstructions for the existence
of multiple eigenvalues for the operator $A(z)$ or for the operator $A^-(z_-)$.

Commuting operators such that the set of eigenvalues of  $A(z)$ does not
intersect  with the set of eigenvalues of $A^-(z_-)$ are called
operators with {\it separated infinities.} Commuting operators have separated
infinities if at least one of the conditions (\ref{r3}), i.e. one of the conditions
$(i)\  m_+n_-\neq m_-n_+,\ \ (ii) \ v^+\neq v^-$, is satisfied.

The set of multiplicities  $\mu_{i_{\pm}}^{\pm}$ of  {\it distinct} eigenvalues
of the operators $A(z), A^-(z_-)$ is called {\it type} of the pair of
commuting operators with separated infinities,
\beq\label{part}
\sum_{i_{\pm}\in I_{\pm}}\mu_{i_{\pm}}^{\pm}=r_{\pm}.
\eeq
Here $I_{\pm}$ are finite sets, which parameterize distinct eigenvalues of
the operators $A(z)$ and $ A^-(z_-)$. Distinct eigenvalues of these operators
correspond to distinct "infinity" \ points on the components
$\G_i$ of the affine spectral curve, defined by the equations $R_i(w,E)=0.$
A simple counting of degree of the divisor of common eigenfunctions
(see details below in the proof of the first statement of Theorem 2.3 )
shows that there is no component $\G_i$, which is compactified by the points
corresponding to eigenvalues of only one of the operators $A(z)$ or $A^-(z_-)$.
The multilicity of an eigenvalue of $A(E)$ is constant on each of the components $\G_i$.
Therefore,

$(i)$ {\it for each index $i_{\pm}$ there is at least one index $j_{\mp}$,
such that} $\mu_{i_{\pm}}^{\pm}=\mu_{j_{\mp}}^{\mp}$.

\noindent
As it seems to the authors, there are no more constraints on the
types of commuting operators with separated infinities, i.e.
for any set of positive integers $\mu_{i_{\pm}}^{\pm}$,
satisfying equation (\ref{part}) and the constraint $(i)$,
there is a pair of commuting operators of the form (\ref{1}) with separated
infinities which has this set as its type.

Below, we prove the statement for the types of the form $(r,r)$.
This is the type of commuting operators of the form (\ref{1})
with separated infinities, which have equal common divisors of positive and
negative orders, $(r=r_+=r_-)$, having the maximum possible scalar rank $\mu=r$.

\begin{lem} Let a pair of irreducible commuting operators
of the form (\ref{1}), where $(r=r_+=r_-)$, have rank $r$. Then
there is a unique Laurent series
\beq \label{r5} v^+(x)=v^+x^{-m_+}+O(x^{-m_++1}))
\eeq
such that the equations
\beq
\label{r6} L\Psi(z)=z^{-rn_+}\Psi(z) ,\ \ A\Psi(z)=v(z^r)\Psi(z)
\eeq
have a solution $\Psi(z)$ of the form
\beq \label{rPsi} \Psi_n(z)=z^{-n}\left(1+\sum_{s=1}^{\infty}
\xi_s(n)z^s\right). \eeq
The space of solutions of equations (\ref{r6}) in the space $\L(z)$
of Laurent series is spanned by the series $\Psi(\epsilon^k z), \epsilon^r=1$.
\end{lem}
Almost identical statement holds for eigenvalues and eigenvectors
of the restriction of the operator  $A$ on the space $\L(z_-)$.

Let us keep the same notation $\G$ for the curve
defined by the equation $R(w,E)=0$, where $R(w,E)$ is a root of degree $r$ from
the characteristic polynomial
$$\det(w-A(E))=R^r(w,E).$$
From Lemma 2.3 and its analog for the operator $A^-(z_-)$ it follows,
that at the infinity we have
\beq
\label{spr} R(w,E)=
\left[\prod_{k=0}^{n_+-1}\left(w-v^+(z_{k})\right)\right]
\left[\prod_{k=0}^{n_--1}\left(w-v^-(z_{k,-})\right)\right],
\eeq
where the products are taken over all the roots of $E^{-1}$ of the orders $n_+$ и
and $n_-$, respectively.

To every point of $\G$ there corresponds $r$-dimensional subspace
of common eigenvectors of the operators $L$ and $A$. Let us fix a basis $\psi^i(Q)$
in this subspace by the conditions
\beq\label{r20}
\psi^i_n(Q)=\delta_{in},\ \ i,n=0,\ldots,r-1. \eeq
For all $n$ the functions $\psi^i_n(Q)$ are meromorphic functions. Their form
in the neighborhood of the infinity can be found with the help
of basic series $\Psi^{\pm}(\epsilon^k z_{\pm})$.
The same as in the continuous case \cite{kr2}, singularities of the vector-function
$\psi_n(Q)=\{\psi_n^i(Q)\}$ in the affine part of the spectral curve
are described by matrix divisors.

In the general position, when the poles of $\psi_n$ are simple, the corresponding
matrix divisor $D=\{\g_s,\a_s\}$ is just a set of distinct points $\g_s$, and a set
of $r$-dimensional vectors $\a_s=\{\a_s^i\}$, defined up to a factor
$\a_s\to \lambda \a_s$. The points $\g_s$ are poles of $\psi_n^i$, and
the parameters $\a_s$ define relations between residues
\beq \label{tu}
\alpha_s^i {\rm res}_{\g_s}\psi_n^j(Q)  =
\alpha_{s}^{j} {\rm res}_{\g_s}\psi_n^i(Q).
\eeq
In \cite{kn1,kr2} the sets $(\gamma,\a)$, were called {\it Tyurin parameters}.
According to \cite{tyur}, in a generic case they
uniquely define a framed stable vector bundle ${\cal E}$ over $\Gamma$ of
rank $r$ and degree $c_1(\det{\cal E})=rg$.

\begin{theo} Let a pair of irreducible commuting operators
with separated infinities satisfies the conditions (\ref{r3}), and has the rank
$r=r_{+}=r_-$. Then:

(1) the spectral curve $\G$, defined by the characteristic equation (\ref{spr})
is irreducible. It is compactified at the infinity by two points $P^{\pm}$,
in the neighborhoods of which local coordinates are $ z_{\pm}=E^{-1/n_{\pm}};$

(2) the vector-function $\psi(Q)$, whose coordinates are common
eigenfunctions
$$L\psi^i_n(Q)=E\psi^i_n(Q),\ \ A\psi^i_n(Q)=w\psi^i_n(Q), \ \ Q=(w,E)\in \G,$$
normalized by the conditions (\ref{r20}), is meromorphic on $\G$;
its matrix divisor outside the punctures $P_{i}^{\pm}$ does not depend on $n$.
In the general position, when the spectral curve is smooth,
the divisor of poles $D=\{\g_s,\a_s\}$ is of degree $gr$, where $g$ is the genus
of $\G$.

(3а) in the neighborhood of $P^+$ the function $\psi^i_{kr+j}(Q)$ has the form
\bea
\psi^i_{kr+j}&=&O(z_+^{-k}), \ \ i<j,\nonumber \\
\psi^i_{kr+i}&=&z_+^{-k}(1+O(z_+)),\nonumber \\
\psi^i_{kr+j}&=&O(z_+^{-k+1}), \ \ i>j. \label{as1} \eea

(3b) in the neighborhood of $P^-$ the function $\psi^i_{kr+j}(Q)$ has the form
\bea
\psi^i_{kr+j}&=&O(z_-^{k+1}), \ \ i<j,\nonumber \\
\psi^i_{kr+j}&=&O(z_-^{k}), \ \ i\geq j. \label{as2} \eea
\end{theo}
The inverse statement holds. For each generic set of points
$\g_s$  and vectors $\alpha_s=(\alpha_{s}^{i}),\ s=1,\ldots,gr,\
i=0,\ldots,r-1$ there is a unique set of functions $\psi_n^i(Q)$ such that:

$(i)$ $\psi_n^i(Q)$ outside the punctures $P^{\pm}$ has at most simple poles at
$\g_s$; their residues at these points satisfy relations (\ref{tu});

$(ii)$ $\psi_n^i(Q)$ at the neighborhoods of the punctures $P^{\pm}$ has
the form (\ref{as1},\ref{as2}).

\begin{theo} Let $\psi^i(Q)=\{\psi_n^i(Q)\}$ be the functions, defined above
by the data $\{\G,P^{\pm}, D=\{\g_s,\alpha_s\}\}$. Then for each function
$f\in \A(\G, P^{\pm})$ there is a unique difference operator $L_f$
(with coefficients independent on $Q$) such, that
$$ L_f\psi^i(Q)=f(Q)\psi^i(Q).$$
If the function $f(Q)$ has poles of orders $n_+$ и $n_-$ at the punctures
$P_i^{\pm}$ then the operator $L_f$ has the form (\ref{1}), where $r=r_+=r_-$.
\end{theo}
{\bf Example. $r=2, g=1$.} Without loss of generality we may assume
that two punctures on the elliptic curve with the periods $(2\omega, 2\omega')$ are
points $\pm z_0$. Let us choose the vectors $\a_1,\a_2$  in the form $\a_s=(a_s,1)$.

From (\ref{as1},\ref{as2}) it follows that $\psi_{2n}^1$ can be represented in the
form
\beq\label{s1}
\psi_{2n}^1(z)=A_n{\s(z-z_0)\s(z-\g_1-\g_2-(2n-1)z_0)\over \s(z-\g_1)\s(z-\g_2)}\
\left[{\s(z+z_0)\over \s(z-z_0)}\right]^n,
\eeq
where $\s(z)=\s(z;2\omega,2\omega')$ is the Weierstrass sigma-function.
Similar expression for $\psi_{2n}^0$ has the form
\beq\label{s2}
\psi_{2n}^0(z)=\left(B_n{\s(z-\g_1-2nz_0)\over \s(z-\g_1)}+
C_n{\s(z-\g_2-2nz_0)\over \s(z-\g_2)}\right)\
\left[{\s(z+z_0)\over \s(z-z_0)}\right]^n,
\eeq
Conditions (\ref{tu}) allow us to express the parameters $B_n, C_n$ in terms of $A_n$
\beq\label{B}
B_n=a_1A_n\ {\s(\g_2+(2n-1)z_0)\s(\g_1-z_0)\over \s(\g_1-\g_2)\s(2nz_0)}
\eeq
\beq\label{C}
C_n=a_2A_n\ {\s(\g_1+(2n-1)z_0)\s(\g_2-z_0)\over \s(\g_2-\g_1)\s(2nz_0)}
\eeq
In the neighborhood of $z_0$ the function $\psi_{2n}^0$ has the form $(z-z_0)^{-n}$.
That implies an explicit formula for $A_n$
\beq\label{A}
A_n={\s(2nz_0)\s(\g_1-\g_2)\over (a_1-a_2)\s^n(2z_0)
\s((2n-1)z_0+\g_2)\s(2n-1)z_0+\g_1)}.
\eeq
In analogous way we find
\beq\label{s11}
\psi_{2n+1}^0(z)=A'_n{\s(z+z_0)\s(z-\g_1-\g_2-(2n+1)z_0)\over \s(z-\g_1)\s(z-\g_2)}\
\left[{\s(z+z_0)\over \s(z-z_0)}\right]^n,
\eeq
\beq\label{s2'}
\psi_{2n+1}^1(z)=\left(B'_n{\s(z-\g_1-2nz_0)\over \s(z-\g_1)}+
C'_n{\s(z-\g_2-2nz_0)\over \s(z-\g_2)}\right)\
\left[{\s(z+z_0)\over \s(z-z_0)}\right]^n,
\eeq
where
\beq\label{B'}
B'_n=a_1^{-1}A'_n\ {\s(\g_2+(2n+1)z_0)\s(\g_1+z_0)\over \s(\g_1-\g_2)\s(2nz_0)}
\eeq
\beq\label{C'}
C'_n=a_2^{-1}A'_n\ {\s(\g_1+(2n+1)z_0)\s(\g_2+z_0)\over \s(\g_2-\g_1)\s(2nz_0)}
\eeq
\beq\label{A'}
A_n'={\s(2nz_0)\s(\g_1-\g_2)\over \s^n(2z_0)}(I_n'-I_n'')^{-1},
\eeq
\beq
I_n'={\s(\g_2+(2n+1)z_0)\s(\g_1+(2n-1)z_0)\s(\g_1+z_0)\over a_1 \s(z_0-\g_1)},
\eeq
\beq
I_n''=
{\s(\g_1+(2n+1)z_0)\s(\g_2+(2n-1)z_0)\s(\g_2+z_0)\over a_2 \s(z_0-\g_2)}.
\eeq
Similar expressions in terms of the Riemann theta-functions
can be found in general case.
\begin{cor} Coefficients of the operators $L_f$, defined by the previous
theorem, are quasi-periodic functions of the variable $n$.
\end{cor}
{\it Remark.} Note that the functions $\psi_{2n}^{(i)}$ and
$\psi_{2n+1}^{(i)}$ in the continuous limit
$$z_0\to 0, \ \ n\to \infty, \ \ nz_0\to x,$$
converge to different functions of the continuous variable $x$.
Probably, that explains why in the classification problem of commuting
differential operators there are no constructions without functional parameters.

\subsection{Rank >1. Glued infinities.}

Now let us consider commuting operators of the form (\ref{1})
having maximum possible rank $l=r_++r_-$. In this case the formal
eigenvalues of the operator $A$ on two infinities $\L_{\pm}$
coincide. An eigenvalue of the operator $A(z)$ may coincide with
an eigenvalue of the operator $A^-(z_-)$ only when $m_+=m_-=m,\
n_+=n_-=n$, i.e. the case of full or partial glued infinities may
occur only in the classification problem of commuting operators of
the form: \beq \label{11a} L=\sum_{i=-Nr_-}^{Nr_+} u_i(n)T^i, \ \
A=\sum_{i=-Mr_-}^{Mr_+}v_i(n)T^i, \ \ (N,M)=1, \eeq We  call {\it
a type} of commuting pair of operators a set $(\mu_i^+|\,\mu_i^-)$,
where the index $i$ is in a set $I$, parameterizing all the distinct
eigenvalues of $A(z)$ и $A^-(z_-)$, and $\mu_i^{\pm}$ are
multiplicities of the corresponding eigenvalues.
Note, that the type of
operators with separated infinities can be identified with the
particular case of the later definition in which $I$ is a union of sets $I_{\pm}$,
which parameterize indices $i_{\pm}$ in (\ref{part}), and all the pairs
have either the form $(\mu_i^+|\,0)$ or $(0|\,\mu_j^-)$.

We conjecture that there are no restrictions on the possible types,
when the numbers $\mu_i^{\pm}>0$ are both non-zero. If the type
contains a pair of the form $(\mu|\, 0)$ then, as before, it
should contain also the pair of the form $(0|\, \mu)$. Complete
solution of the classification problem of commuting difference
operator requires a construction of commuting operators of all the
possible types. We leave the problem open, and consider below a
construction of commuting operators of the type that contains only
one pair $(r_+|\,r_-)$. As it was mentioned above, that is the
case of commuting operators of the maximum possible rank $l=r_++r_-$.

\medskip
\noindent
{\bf Direct spectral problem.}
From the construction of the formal eigenfunctions at the infinity
it follows, that the necessary conditions for the case of the maximum
rank are  equations $v_i^+=v_j^-=v$, i.e. the operators have the from (\ref{11a}),
and their leading coefficients in the gauge $u_{Nr_+}=1$ have the from
\beq\label{71b} u_{-Nr_-}=h^{-1}(n-Nr_-)h(n), \ \
v_{Mr_+}=v,\ \ v_{-Mr_-}=vh^{-1}(n-Mr_-)h(n). \eeq
As above,  we keep the same notation $\G$ for the curve
defined by the equation $R(w,E)=0$, where $R(w,E)$ is a root of degree $l$ from
the characteristic polynomial
$$\det(w-A(E))=R^{l}(w,E).$$
At the infinity we have the expansion
\beq
\label{spr1} R(w,E)=
\left[\prod_{k=0}^{N-1}\left(w-v(z_{k})\right)\right],
\eeq
where the product is taken over all the roots of $E^{-1}$ of degree $N$.
Hence, in the case of the maximum rank the spectral curve is compactified
at the infinity by {\it one} smooth point, and as a corollary, it is irreducible.

To every point of $\G$ there corresponds $l$-dimensional space of common
eigenfunctions of the operators $L$ and $A$. Let us fix the basis $\psi^i(Q)$ of
this space by the conditions
\beq\label{r200}
\psi^i_n(Q)=\delta_{in},\ \ -r_-\leq i,n<r_+.
\eeq
Note, that the choice of the interval for the parameters $i,n$, used for
the normalization is crucial for closed description of analytical properties
of the basis functions at the infinity.
\begin{theo} For a generic pair of commuting rank $l$ operators, the common
eigenfunctions $\psi_n^i$, normalized by  (\ref{r200}), satisfy
the following conditions:

$1^0.$ On the affine part of the spectral curve $\G$ the functions
$\psi_n^i$ have $gl$  poles $\g_s$ independent of $n$, at which
the equations \beq \label{tu1a} \alpha_s^j \,{\rm
res}_{\g_s}\psi_n^i(Q) =\alpha_s^i \,{\rm res}_{\g_s}\psi_n^j(Q).
\eeq hold.

$2^0.$ In the neighborhood of the "infinity" \ point $P_0$ the vector-row
$\psi_n=\{\psi_n^i\}$ has the form
\beq\label{ps22}
\psi_n=\left(\sum_{s=0}^{\infty}\xi_s(n)z^{s}\right)\Psi_0(n,z),\ \ \ z^{-n}=E.
\eeq
Here $\xi_s(n)=\{\xi_s^{i}(n)\}$ are vector-rows,
\beq\label{ps222}
\xi_0^i=\delta^i_0\, ;
\eeq
$\Psi_0(n,z)$ is the Wronsky matrix
$\Psi_0^{j\,i}(n,z)=\phi_{n+j}^i(z),\, -r_-\leq i,j<r_+,$
defined by the basis $\phi^i$ of the solutions of the equation
\beq\label{ps42}
\phi_{n+r_+}+\sum_{i=-r_-}^{r_+-1}f_i^{\,0}(n)\phi_{n+i}=z^{-1}\phi_n,\ \
\eeq
whose coefficients $f_i^{\,0}(n)$ are polynomial functions
of the coefficients of the operator $L$.
The solutions $\phi^i$ are normalized by the equation
\beq\label{ps322}\Psi_0(0,z)=1.
\eeq

\end{theo}

\noindent
{\it Proof.} Let $\Psi(n,Q), \ Q\in \G,$
be the Wronsky matrix $\Psi^{j\,i}(n,Q)=\psi_{n+j}^i(Q),\  -r_-\leq i,j<r_+\, $.
In the neighborhood of $P_0$, where $z=E^{-1/n}$ is a local coordinate,
it can be considered as a function of the variable $z$, i.e.
$\Psi(n,z)$. Let us define polynomials $\phi_n^i$ of the variable $z^{-1}$,
by fixing their asymptotic behavior  as $z\to 0$.
\begin{lem}  In a generic case there are unique functions $\phi_n^i(z)$,
that are holomorphic on the extended complex $z$-plane everywhere except at
$z=0$, and such that in the neighborhood of $z=0$ the vector-row
$\phi_n(z)=(\phi_n^i(z))$ has the form
\beq\label{pol}
\phi_n(z)=r_n(z)\Psi(n,z),
\eeq
where $r_n(z)$ is a vector-row holomorphic in the neighborhood of $z=0$, and such,
that its evaluation at $z=0$ equals
\beq\label{pol1}
r_n^i(0)=\delta^i_0.
\eeq
\end{lem}
The problem of construction of $\phi_n(z)$ is a standard Riemann problem.
Let us choose a small neighborhood of $z=0$. Then we define
vector-functions $\phi_n$ и $r_n$ as holomorphic outside and inside
the neighborhood, respectively, and such that on the boundary of the neighborhood
the equation (\ref{pol}) holds. If the change of the argument of the determinant
of $\Psi(n,z)$ along the contour equals zero, then the Riemann has
a unique solution if we fix a value of the vector-function at some point.
Therefore, for the proof of the Lemma it is enough to show that the determinant
of $\Psi(n,z)$ is holomorphic in the neighborhood of $z=0$  and in general
does not vanish at $z=0$. The later statement follows from:

\begin{lem} In the neighborhood of the infinity
the matrix function
\beq\label{W}
\X(n,Q)=\Psi(n+1,Q)\Psi^{-1}(n,Q)
\eeq
has the form
\beq\label{ww4}
\X(n,z)=\left(\begin{array}{ccccc}
0& 1&0&\cdots &0\\
0& 0&1&\cdots &0\\
\cdot&\cdot&\cdot&\cdot&\cdot\\
0& 0&0&\cdots &1\\
\chi_{-r_-}\,(n,z)& \chi_{-r_-+1}\,(n,z)&\chi_{r_-+2}\,(n,z)&\cdots &\chi_{r_+-1}\,(n,z)
\end{array}\right)
\eeq
\beq\label{ww5}
\chi_i(n,z)=z^{-1}\delta_{i,\,0}-f_i(n,z),
\eeq
where $f_i(n,z)$ are regular series of the variable $z$.
\end{lem}
{\it Proof.} The matrix $\X(n,Q)$ does not depend on the choice of
the basis functions $\psi_n^i$. Therefore, the asymptotic behavior
of $\X(0,Q)$ in the neighborhood of the infinity, can be found with the help
of the formal solutions defined in the subsection 2.2. Equations
(\ref{psi},\ref{psi-}) imply that the Wronsky matrix
$\Psi_{\infty}(0,z)$, defined by the formal solutions, has the following block form
\beq\label{ww}
\Psi_{\infty}(0,z)=\left(\begin{array}{cc}z^{-1}A(z)& B(z)\\C(z)&D(z)\end{array}\right),
\eeq
where $A(z),D(z)$ are matrices of dimensions  $(r_-\times r_-)$
and  $(r_+\times r_+)$, respectively. All the matrices $A,B,C,D$ are regular
series in the variable $z$. The leading term of $D(z)$
is a lower triangular matrix with 1 on the diagonal. The leading term
of the series $A(z)$ is an upper triangular matrix.
Hence, the inverse matrix has the form
\beq\label{ww1}
\Psi^{-1}_{\infty}(0,z)=\left(\begin{array}{cc}zA_1(z)& zB_1(z)\\zC_1(z)&D_1(z)\end{array}
\right)
\eeq
The leading terms of the regular series $A_1,B_1,C_1,D_1$ equal
\begin{eqnarray}
A_1(0)=&A^{-1}(0),\ B_1(0)=&-A^{-1}(0)B(0)D^{-1}(0),\nonumber\\
D_1(0)=&D^{-1}(0),\ C_1(0)=&-D^{-1}(0)C(0)A^{-1}(0). \label{ww2}
\end{eqnarray}
The series $\psi_{r_+}^i$ have the form
$z^{-1}\delta_0^i+f^i(z)$, where $f^i$
are regular. Therefore, the last row of the matrix $\X(0)$, which is equal to
$\psi_r\Psi_{\infty}^{-1}(0)$, has the form
\beq\label{ww3}
\X^{r_+-1,\,i}(0)=z^{-1}\delta_0^i+f^i(z).
\eeq
In the construction of the formal solutions index $n$ can be replaced by
$n-n_0$. That shift does not change $\X(n)$. Therefore, the last row of the
matrix $\X(n_0)$ has the same structure as for $\X(0)$, and
the Lemma is proven.

Normalization (\ref{r200}) is equivalent to the equation $\Psi(0,z)\equiv 1$.
From (\ref{ww4}) it follows that
$$\det \Psi(n,z)=\prod_{m=0}^{n-1}\det \X(n,z)=(-1)^n\prod_{m=0}^{n-1}f_{-r_-}(n,z).$$
Hence the determinant is holomorphic in the neighborhood of $z=0$, and in general
does not vanish at $z=0$. That completes the proof of
the Lemma 2.4.

Note, that by their construction the functions $\phi_n^i$ are entire functions
of the variable $z^{-1}$. The matrix $\Psi(n,z)$
is meromorphic in the neighborhood of $z=0$. Hence, $\phi_n^i$ at $z=0$
has the pole of finite order and, therefore is a polynomial function of
$z^{-1}$.

Let $\Psi_0(n,z)$ be the Wronsky matrix of the functions $\phi_n^i(z)$, i.e.
$\Psi^{j,\,i}(n,z)=\phi_{n+j}^i(z)$.
\begin{lem} In the neighborhood of \, $z=0$ the matrix function $\Psi_0$ has the form
\beq\label{w11}
\Psi_0(n,z)=R(n,z)\Psi(n,z),
\eeq
where $R(n,z)$ is a matrix function holomorphic at $z=0$,
and such that $R(n,0)$ has the block form
\beq\label{www}
R(n,0)=\left(\begin{array}{cc} R_-&0\\0&R_+\end{array}\right)
\eeq
where $R_+$ is a lower-triangular, and  $R_{-}$ is an upper triangular
$(r_{\pm}\times r_{\pm})$ matrices with 1 on the diagonals.
\end{lem}
{\it Proof.} From the definition of $\phi_n$ it follows that $j$-th row $R_j$
of the matrix $R$ for $j> 0$ equals
\beq\label{www1}
R_j(n,z)=r_{n+j}(n,z)\prod_{i=0}^{j-1}\X(n+i,z)
\eeq
Equations (\ref{pol1}, \ref{ww5}) imply that $R_j$ is regular. Moreover,
coordinates $R_j^{\,i}(n,0)$ of the vector $R_j(n,0)$ might be different from
zero only for $0\leq i\leq j$. At the same time $R_j^j(n,0)=1.$
That proves that $R_+$ is an upper triangular matrix with 1 on the diagonal.
The analogous statement for $R_-$ can be proven by similar arguments
if one replaces $\X$ by the inverse matrix $\X^{-1}$ which has the form
\beq\label{www2}
\X^{-1}=\chi_{-r_-}^{-1}\left(\begin{array}{cccccc}
\chi_{-r_-+1}& \chi_{-r_-+2}&\chi_{-r+3}&\cdots& \chi_{r_+-1}&1\\
1& 0&0&\cdots &0&0\\
0& 1&0&\cdots &0&0\\
\cdot&\cdot&\cdot&\cdot&\cdot&\cdot\\
0& 0&0&\cdots &1&0\\
\end{array}\right)
\eeq
By the definition $\Psi(0,z)=1$. Entries of $\Psi_0$ are polynomials in the
variable $z^{-1}$. Therefore, (\ref{w11}) implies that $\Psi_0(0,z)$
is a constant matrix, equal to $R(0,0)$. By induction in $j$, it is easy to show
that $\Psi_0(0,z)$ satisfies the normalization (\ref{ps322}), i.e.
$\Psi_0(0,z)=1.$

Let us prove that the functions $\phi_n^i$ satisfy an equation of the form (\ref{ps42}).
Consider the matix
\beq\label{as}
\Psi_0(n+1,z)\Psi_0^{-1}(n,z)=\X_0(n,z)=R(n+1,z)\X(n,z)R(n,z)^{-1}\, .
\eeq
From (\ref{ww5}) and the structure of $R(n,0)$, it follows that
the only entry of the last row in $\X_0$, having pole at $z=0$
is $\X_0^{r_+-1,\,0}.$ A coefficient at the singular term
$z^{-1}$ of its expansion equals 1. The matrix $\X_0$ is regular for
all $z\neq 0$. Therefore, it is a polynomial function of $z^{-1}$.
Hence the last row of $\X_0$ has the desired form.

In order to complete the proof of the second statement of the Theorem it is enough
to invert equation (\ref{w11}). We have
\beq\label{as11}
\Psi=R^{-1}\Psi_0.
\eeq
This equation for the row with the index $j=0$ gives (\ref{ps22}),
where the first factor is the Tailor expansion of
the corresponding row of the matrix $R^{-1}(n,z)$.

Using the asymptotic of $\psi_n$ at the infinity  one can find
the degree of the matrix divisor of its poles
using arguments identical to that in the continuous case \cite{kr2}.

The Theorem 2.3. assigns for the generic
pair of commuting operator of the form (\ref{11a}), having rank $l=(r_++r_-)$ :
a smooth algebraic curve $\G$ with a puncture $P_0$; a set of Tyurin parameters
and a set of $l-1$ functions $f_i^{\,0}(n)$ of the discrete variable $n$
\beq\label{soot}
L,A\longmapsto \{\G,(\g,\a),f_i^{\,0}(n)\}.
\eeq

\medskip
\noindent
{\bf Inverse spectral problem.} Let $\G$ be a smooth algebraic curve
with fixed local coordinate $z$ at the neighborhood of puncture $P_0$.
Let us fix a set of function $f_i^{\,0}(n),\ \ r_-\leq i<r_+$. They define
the Wronsky matrix $\Psi_0,\ \Psi_0^{j,\,i}=\phi_{n+j}^i, \ r_-\leq j<r_+$,
corresponding to the solutions of the degree $l=r_++r_-$ difference equation
\beq\label{op}
\sum_{i=r_-}^{r_+}f_i^{\,0}(n)\phi_{n+i}=z^{-1}\phi_n,\ \ f^0_{r_+}=1,
\eeq
normalized by the conditions
\beq\label{nor22}
\phi_{n}^i=\delta^i_0,\ \ r_-\leq i<r_+.
\eeq
\begin{theo} (\cite{kn})
For a generic set of Tyurin parameters of degree $lg$
and rank $l$, i.e. for a generic set of $lg$ points $\g_s$ and a set of
projective $l$-dimensional vectors $\a_s=(\a_s^i),\ r_-\leq i\leq r_+$
there is a unique vector-function $\psi_n(Q)$ with coordinates, which
outside the puncture $P_0$ have at most simple poles at $\g_s$. Their
residues satisfy relations (\ref{tu1a}). In the neighborhood of $P_0$
the vector-row $\psi_n$ has the form
\beq\label{ps22a}
\psi_n=\left(\sum_{s=0}^{\infty}\xi_s(n)z^s\right)\Psi_0(n,z),
\ \xi_0^i=\delta^i_0\, .
\eeq
Let $f\in \A(\G,P_0)$ be a meromorphic function on $\G$
with the only pole at $P_0$ of order $N$. Then,
there is a unique operator $L_f$ of the form
\beq\label{N}
L_f=\sum_{i=-Nr_-}^{Nr_+}u_i(n)T^i,\ \ u_{Nr_+}=1,
\eeq
such that
\beq\label{lf}
L_f\psi(Q)=f(Q)\psi(Q)\, .
\eeq
\end{theo}
A proof of the theorem is standard. Equation (\ref{ps22a}) means that
$\psi$ is a solution of the Riemann problem on $\G$, in which $\Psi_0$ is
the transition function in the neighborhood of the puncture. In the general position
the existence and uniqueness of the solution of this problem follows from the
results of \cite{rodin, koppelman}, or simply from the Riemann-Roch theorem
for vector bundles (see details in \cite{kr2}). The same results imply
the second statement of the theorem.

\begin{theo} A commutative ring $\A$ of operators of the form (\ref{N})
of maximum rank $l$ is isomorphic to the ring $\A(\G,P_0)$ of meromorphic functions
on an algebraic curve $\G$ with the only pole at the puncture $P_0$.
In the general case the isomorphism $\A(\G,P_0)=\A$ is defined by
equation (\ref{lf}), where the Baker-Akhiezer function corresponds to a set
of the Tyurin parameters $(\g,\a)$.
\end{theo}

\subsection{Discrete dynamics of the Tyrin parameters}

In this section discrete equations for the Tyurin parameters
corresponding to commuting operators of maximum rank $l$ are derived.
According to Theorem 2.6, a smooth algebraic curve $\G$
with a puncture $P_0$, a generic set of the Tyurin parameters $(\g,\a)$ and
coefficients $f_i^{\,0}(n)$ of difference equation (\ref{op}) define
the vector-function $\psi_n(Q)$. The corresponding Wronsky matrix
is denoted by $\Psi(n,Q)$. The matrix function  $\X(n,Q)$,
given by (\ref{W}), has the asymptotic (\ref{ww4}, \ref{ww5}).
Let
\beq\label{F1}
f_i(n)=f_i(n,0)
\eeq
be the evaluations of the regular series $f_i(n,z)$ in (\ref{ww5}) at the puncture
$z=0$. These functions of the discrete variable $n$ can be expressed
in terms of the initial parameters $f_i^{\,0}(n)$, and the first
coefficients $\xi_1(n)$ of expansion (\ref{ps22a}) for $\psi_n$.
The corresponding formulae are far from being effective, because
the coefficients $\xi_1(n)$ depend on the initial data
$(\g_s,\ \a_s, \ f_i^{\,0}(n)$ through the Riemann problem.
At the same time, as it will be shown below, there is no need for explicit formulae
for $f_i(n)$. They can be chosen as a part of independent parameters defining
commuting operators.

For  $n\neq 0$ let us denote zeros of $\det \Psi(n,Q)$ by $\g_s(n)$.
In general position, when they are simple their number equals $gl$.
Let $\a_s(n)$ be the corresponding left zero-vector
\beq\label{F2}
\a_s(n)\Psi(n,\g_s(n))=0\, .
\eeq
For $n=0$, we set
\beq\label{p3}
\g_s(0)=\g_s,\ \ \a_s(0)=\a_s.
\eeq
From the definition (\ref{W}) it follows:
\begin{lem}
The matrix function $\X(n,Q)$ has simple poles at the points $\g_s(n)$
and satisfies the relations
\beq\label{ps22b}
\a_s^j(n)\ {\rm res}_{\g_s(n)}\X^{m,i}(n,Q)=
\a_s^i(n)\ {\rm res}_{\g_s(n)}\X^{m,j}(n,Q).
\eeq
The points $\g_s(n+1)$ are zeros of $\X(n,Q)$,
\beq\label{p4}
\det \X(n,\g_s(n+1))=0.
\eeq
The vector $\a_s(n+1)$  is the left eigenvector of the matrix $\X(n,\g_s(n+1))$,
\beq\label{p5}
\a_s(n+1)\X(n,\g_s(n+1))=0.
\eeq
\end{lem}
Simple dimension counting implies the following statement.
\begin{lem} Let $\G$ be a smooth algebraic curve with fixed
local coordinate с  $z(Q)$ in the neighborhood of a puncture
$P_0$. Then for a generic set of data $ (\g_s(n),\ \a_s(n), \ f_i(n))$ there
is a unique meromorphic matrix function $\X(n,Q),\ Q\in\G$,
with at most simple poles at the points$P_0,\g_s$, and such that:

$(i)$ the Laurent expansion of $\X(n,Q)$ at $P_0$
has the form (\ref{ww4},\ref{ww5}), where the regular series $f_i(n,z)$
satisfy (\ref{F1});

$(ii)$ the residues of $\X(n,Q)$ at  $\g_s$ satisfy the relations (\ref{ps22b}).
\end{lem}
Equations (\ref{p4}, \ref{p5}) can be regarded as equations
that define  $(\g_s(n+1),\a_s(n+1))$ through the matrix $\X(n,Q)$.
The later is uniquely defined by $(\g_s(n), \a_s(n), f_i(n))$. Hence:
\begin{cor} The functions $f_i(n)$  and the Tyurin parameters $(\g,\a)$,
which define the initial condition (\ref{p3}) of the corresponding
discrete dynamical system, are the full set of data, which parameterize
commuting operators with the given spectral curve.
\end{cor}

\bigskip
\noindent
{\bf Example $g=1, \ l=2$}. Let us consider a pair
of commuting operators
\beq\label{pr}
L=\sum_{i=-2}^2u_i(n)T^i, \ \ A=\sum_{i=-3}^3v_i(n)T^i
\eeq
of the maximum rank $l=2$. Their spectral curve $\G$
is an elliptic curve. Let $2\omega,2\omega'$ be periods of $\G$.
Without loss of generality we identify  the puncture $P_0$ with the point
$z=0$ in the fundamental domain of $z$-plane.

The operators (\ref{pr}) are uniquely defined by Tyurin parameters, and the
parameters $f_i(n),\ i=-1,0$, which are denoted below by
\beq\label{pr2}
f_{-1}=c_{n+1}, \ \ f_0=v_{n+1}.
\eeq
Our next goal is to find explicit formulae
for the coefficients of commuting operators
(\ref{pr}) with the help of equations of the discrete {\it dynamics}
of Tyurin parameters. For brevity, let us introduce the notations
\beq\label{pr3}
\g_n^1=\g_1(n),\ \ \ \g_n^2=\g_2(n)\,.
\eeq
For the vectors $\a_s^i,\  \ i=-1,0$, which are defined up to a factor,
we fix the normalization under which $\a_s(n)$ are two-dimensional row-vectors with
the coordinates
\beq\label{pr4}
\a_1(n)=(a_n^1,1),\ \ \a_2(n)=(a_n^2,1).
\eeq
According to Lemma 2.8, the data $\g_n^{1,2},\ a_n^{1,2},\ c_{n+1},\ v_{n+1}$
uniquely define the matrix $\X_n^{ji}=\X_n^{ji}(n,z),\ \ i,j=-1,0.$
Let us find its explicit form in terms of the standard Weierstrass functions.
By definition, $\X_n$ has the form
\beq\label{pr5}
\X_n=\left(\begin{array}{cc} 0&1\\ \chi^1_n(z)&\chi^2_n(z)\end{array}\right)
\eeq
An elliptic function $\chi_n^1(z)$ has poles at $\g_n^{1,2}$ and equals $-c_{n+1}$ at
$z=0$.
Therefore, it can be written as
\beq\label{p8}
\chi_n^1=-c_{n+1}+A_1\left(\zeta(z-\g_n^1)+\zeta(\g_n^1)\right)+
B_1\left(\zeta(z-\g_n^2)+\zeta(\g_n^2)\right),
\eeq
where $\zeta(z)$ is the Weierstrass $\zeta$-function.

The function $\chi_n^2$ in the neighborhood of the puncture $z=0$ has the form
$\chi^2=z^{-1}-v_{n+1}+O(z)$, i.e.
\beq\label{p9}
\chi_n^2=-v_{n+1}+\zeta(z)+A_2\left(\zeta(z-\g_n^1)+\zeta(\g_n^1)\right)+
B_2\left(\zeta(z-\g_n^2)+\zeta(\g_n^2)\right).
\eeq
The sum of residues of an elliptic function equals zero. Hence,
\beq\label{p10}
A_1+B_1=0,\ \ A_2+B_2=-1.
\eeq
At the same time
\beq\label{p11}
A_1=a_n^1A_2,\ \ B_1=a_n^2B_2.
\eeq
From (\ref{p10},\ref{p11}) it follows that
\begin{eqnarray}
\chi_n^1&=&-c_{n+1}+{a_n^1a_n^2\over a_n^1-a_n^2}
\left(\zeta(z-\g_n^1)-\zeta(z-\g_n^2)+\zeta(\g_n^1)-\zeta(\g_n^2)\right),\label{p12}\\
\chi_n^2&=&\zeta(z)-v_{n+1}+
{a_n^2\over a_n^1-a_n^2}\left(\zeta(z-\g_n^1)+\zeta(\g_n^1)\right)+\nonumber\\
{}&{}&+{a_n^1\over a_n^2-a_n^1}\left(\zeta(z-\g_n^2)+\zeta(\g_n^2)\right)\,.
\label{p13}
\end{eqnarray}
According to Lemma 2.7, the points $\g_{n+1}^s$ are defined by the equation\\
$\det \X_n(\g_{n+1}^s)=\chi^1(\g_{n+1}^s)=0$. Hence,
\beq\label{p16}
c_{n+1}={a_n^1a_n^2\over a_n^1-a_n^2}
\left(\zeta(\g_{n+1}^s-\g_n^1)-\zeta(\g_{n+1}^s-\g_n^2)+
\zeta(\g_n^1)-\zeta(\g_n^2)\right)\,.
\eeq
Note, that the sum $\g_n^1+\g_n^2=2c$  does not depend on  $n$, because $\g_n^s$ are poles
and $\g_{n+1}^s$ are zeros of an elliptic function.
Therefore,
\beq\label{p17}
\g_n^1=\g_n+c,\ \ \g_n^2=-\g_n+c,\ \ \ c=const.
\eeq
Using (\ref{p17}), equations (\ref{p12},\ref{p13}) can be rewritten as
\begin{eqnarray}
\chi_n^1&=&{a_n^1a_n^2\over a_n^1-a_n^2}
\left[\zeta(z-\g_n-c)-\zeta(z+\g_n-c)-\zeta(\g_{n+1}-\g_n)-
\zeta(\g_{n+1}+\g_n)\right]
\label{p18}\\
\chi_n^2&=&-v_{n+1}+\zeta(z)+
{a_n^2\over a_n^1-a_n^2}\left(\zeta(z-\g_n-c)+\zeta(\g_n+c)\right)+ \nonumber\\
{}&{}&+{a_n^1\over a_n^2-a_n^1}\left(\zeta(z+\g_n-c)-\zeta(\g_n-c)\right)\,.
\label{p19}
\end{eqnarray}
If the functions $v_n$ and $\g_n$ of the discrete variable $n$ are chosen as independent
parameters, then equation (\ref{p16})
\beq\label{p22}
c_{n+1}={a_n^1a_n^2\over a_n^1-a_n^2}
\left(\zeta(\g_{n+1}-\g_n)-\zeta(\g_{n+1}+\g_n)+\zeta(\g_n+c)+\zeta(\g_n-c)\right)\,
\eeq
can be regarded as the definition of the variables $c_{n+1}$.

From (\ref{p5}) it follows that
\beq\label{p6}
a_s(n+1)=-\chi^2_n(\g_s(n+1)).
\eeq
Equation (\ref{p19}), implies recurrent relations defining $a_{n+1}^{1,2}$:
\bea
a_{n+1}^1&=&v_{n+1}-\zeta(\g_{n+1}+c)-
{a_n^2\over a_n^1-a_n^2}\left(\zeta(\g_{n+1}-\g_n)+\zeta(\g_n+c)\right)\nonumber\\
{}&{}&-
{a_n^1\over a_n^2-a_n^1}\left(\zeta(\g_{n+1}+\g_n)-\zeta(\g_n-c)\right),\label{p23}\\
a_{n+1}^2&=&
v_{n+1}+\zeta(\g_{n+1}-c)+
{a_n^2\over a_n^1-a_n^2}\left(\zeta(\g_{n+1}+\g_n)-\zeta(\g_n+c)\right)\nonumber\\
{}&{}&-{a_n^1\over a_n^1-a_n^2}\left(\zeta(\g_{n+1}-\g_n)+\zeta(\g_n-c)\right).\label{p24}
\eea
As shown above, the arbitrary functions $\g_n,v_n$ and a certain constant $c$
define the matrix function $\X_n$, and as a corollary,
define the coefficients of commuting rank $2$ operators, corresponding
to an elliptic curve. Each operator corresponds to an elliptic functions
with the pole at the puncture $z=0$. The simplest operator
$L_4$ is of order $4$ and corresponds to the Weierstrass function
$\wp(z)=z^{-2}+0(z^2)$.

Coefficients of this operator are coefficients of the expansion
of $\wp(z)\psi_n\Psi_n^{-1}$
in the basis $\psi_{n+i}\Psi_{n}^{-1}, 2\leq i\leq 2$. The expansion
is determined by singular parts of the Laurent series at $z=0$.
Let $\tilde\psi_m$ be polynomials in $k=z^{-1}$, such that
$\psi_m\Psi_n^{-1}=\tilde\psi_m+0(k^{-1})$. Then
\beq\label{p25}
\tilde \psi_{n+2}=(-c_{n+1},k-v_{n+1}),\ \ \tilde\psi_{n+1}=(0,1),\ \ \tilde\psi_n=
(1,0).
\eeq
The polynomial $\tilde\psi_{n-1}$ can be found with the help of the equation
$\Psi_{n-1}=\X_{n-1}^{-1}\Psi_n$,
\bea\label{p26}
\X_{n}^{-1}={1\over \chi_n^1}\left(\begin{array}{cc} -\chi_n^2,& 1\\
\chi_n^1,& 0\end{array}\right)\,.
\eea
Let $\xi_n^{ij}$ be coefficients of the expansions
\bea
\chi_n^1&=&-c_{n+1}\left(1+\xi_n^{11}z+\xi_n^{12}z^2+\cdots\right)\label{p27}\\
\chi_n^2&=&k-v_{n+1}+\xi_n^{21}z+\cdots\label{p28}
\eea
Then
\beq\label{p29}
\tilde\psi_{n-1}=c_n^{-1}\left(k-v_n-\xi_{n-1}^{11}\ ,-1\right)\,.
\eeq
In a similar way $\tilde\psi_{n-2}$ can be found.
After a long but straightforward calculation we find
\beq\label{30}
L_4=L_2^2-\left(\xi_{n-1}^{11}+\xi_{n-2}^{11}\right)T+c_n\left(\xi_{n-1}^{11}+\xi_{n-2}^{11}\right)T^{-1}
+u_n\,,
\eeq
where $L_2$ is the discrete Schr\"odinger operator with the coefficients
\beq\label{31}
L_2=T+v_n+c_nT^{-1},
\eeq
and the function $u_n$ is given by the formula
\beq\label{32}
u_n=v_n\left(\xi_{n-1}^{11}-\xi_{n-2}^{11}\right)+\xi_{n-1}^{12}+\xi_{n-2}^{12}-
\left(\xi_{n-2}^{11}\right)^2-\left(\xi_{n-1}^{21}+\xi_{n-2}^{21}\right)\,.
\eeq

\medskip
\noindent
{\bf Symmetric case}. Let the constant $c$  in (\ref{p17}) be zero, $c=0$.
Then $\chi_n^1$ is an even function of $z$. Hence, $\xi_{n}^{11}=0$.
Equation (\ref{p18}) implies
\beq\label{33}
\xi_{n}^{12}=-{a_n^1a_n^2\over (a_n^1-a_n^2)}\ {\wp'(\g_n)\over c_{n+1}}\,.
\eeq
Using (\ref{p22}), we obtain
\beq\label{331}
\xi_n^{12}=
{\wp'(\g_n)\over \zeta(\g_{n+1}+\g_n)-\zeta(\g_{n+1}-\g_n)-2\zeta(\g_n)}=\wp(\g_{n})-
\wp(\g_{n+1})\,.
\eeq
(The last equation can be obtained from
the addition formulae for the Weierstrass $\zeta$-function,
or by direct comparison of the poles and the residues of both sides
of the equation.)
From  (\ref{p19}) it follows that
\beq\label{34}
\xi_n^{21}=\wp(\g_n)\,.
\eeq
Substitution of the last two formulae into (\ref{32}) implies that the operator
$L_4$ in the symmetric case equals
\beq\label{35}
L_4=L_2^2-\wp(\g_n)-\wp(\g_{n-1}).
\eeq
In the symmetric case the formulae for
the coefficients of $L_2$ can by simplified. Let $F(u,v)$ be the elliptic
function
\beq\label{kn10}
F(u,v)=\zeta(u+v)-\zeta(u-v)-2\zeta(v)={\wp'(v)\over \wp(v)-\wp(u)}\,.
\eeq
Then, the formulae (\ref{p22}-\ref{p24}) for the symmetric case
$c=0$ can be represented in the form
\beq\label{kn7}
c_{n+1}=-{a_n^1a_n^2\over a_n^1-a_n^2} F(\g_{n+1},\g_n),
\eeq
\beq\label{kn8}
a_{n+1}^1=v_{n+1}+{1\over 2}\left(F(\g_n,\g_{n+1})+
{a_n^1+a_n^2\over a_n^1-a_n^2}F(\g_{n+1},\g_n)\right),
\eeq
\beq\label{kn9}
a_{n+1}^2=v_{n+1}-{1\over 2}\left(F(\g_n,\g_{n+1})-
{a_n^1+a_n^2\over a_n^1-a_n^2}F(\g_{n+1},\g_n)\right)\,.
\eeq
The last two equations are equivalent to
\bea
a_{n+1}^1-a_{n+1}^2&=&F(\g_n,\g_{n+1}),\label{kn11}\\
a_{n+1}^1+a_{n+1}^2&=&2v_{n+1}+{a_n^1+a_n^2\over a_n^1-a_n^2}\,F(\g_{n+1},\g_{n}).
\label{kn12}
\eea
Let us define $s_n$ by the formula
\beq\label{kn13}
s_n=-{a_n^1+a_n^2\over a_n^1-a_n^2}\,.
\eeq
Then
\beq\label{kn14}
a_n^1+a_n^2=-s_n F(\g_{n-1},\g_{n}), \ \ {a_n^1a_n^2\over a_n^1-a_n^2}=
-{1\over 4}\left(s_n^2-1\right)F(\g_{n-1},\g_{n})
\eeq
and equations (\ref{kn7}, \ref{kn12}) can be rewritten as
\bea
4c_{n+1}&=&(s_n^2-1)F(\g_{n+1},\g_n)F(\g_{n-1},\g_n),\label{kn15a}\\
2v_{n+1}&=&s_n F(\g_{n+1},\g_n)-s_{n+1} F(\g_{n},\g_{n+1})\label{kn16a}.
\eea
The last equation shows, that in  the symmetric case
the variables $\g_n, s_n$ can be chosen as independent parameters.
Then, equations (\ref{31},\ref{35},\ref{kn15a},\ref{kn16a}) give
explicit expressions for the coefficients of the operator
$L_4$, which were presented in the introduction. In the similar way
one can find coefficients of the operator $A_6$ .

\section{High rank solutions of the  $2D$ Toda lattice equations.}

A key element of the algebro-geometric integration theory of
non-linear equations is a construction of the {\it multiparametric}
Baker-Akhiezer functions. These functions are defined
by their analytic properties on a corresponding algebraic curve.
Below we define the multi-parametric Baker-Akhiezer vector-functions.
They  are deformations of the eigenfunctions of commuting operators
of an arbitrary rank. The definitions are different in the cases of operators
with separated or glued infinities.

\bigskip
\subsection{Separated infinities.}
As it was shown above, operators with separated infinities are defined by their
algebro-geometric data: a spectral curve with two punctures
and a set of Tyurin parameters. There no functional parameters in their
construction. Functional parameters do appear in construction of the
corresponding solutions of the $2D$ Toda lattice equations.
It is necessary to stress that these functional parameters
are functions of continuous variables, but not a discrete one,
as in the construction
of commuting operators with glued infinities. The functional parameters
define {\it germ} functions $\Psi_{\pm}(t^{\pm},z)$,
depending on the corresponding half of the times
of the hierarchy of the $2D$ Toda, and are entire functions of
the variable $z^{-1}$.

Consider functions $\Psi_{\pm}(z)$ of the variable $z^{-1}$, such that
\beq\label{arg}
\oint_{|z|=\varepsilon} d\left(\ln \det \Psi_{\pm}\right)=0.
\eeq
\begin{lem}
Let $\G$ be a smooth genus $g$ algebraic curve
with fixed local coordinates $z_{\pm}$ in the neighborhoods of two punctures
$P_{\pm}$. Then for a generic set of the Tyurin parameters
of degree $rg$ and rank $r$ there is a unique vector-function
$\psi_n(Q)$ such that:

$(i)$ its coordinates $\psi_n^i, \ i=0,\ldots, r-1,$ outside the punctures
$P^{\pm}$ have at most simple poles at the points $\g_s$, where
the relations (\ref{tu}) hold;

$(ii)$ in the neighborhoods of the punctures $P^{\pm}$ the function
$\psi_n$ has the form
\beq\label{as4}
\psi_{kr+j}=z_{\pm}^{{\mp}k}R_{\pm}(kr+j,z_{\pm})
\Psi_{\pm}(z_{\pm})\ ,
\eeq
where $R_{\pm}(n,z)$ are holomorphic in the neighborhood of $z=0$
row-vectors with values at $z=0$ satisfying the normalization conditions
\beq\label{as5}
R_-^{\,i}(kr+j,0)=0, \ i\leq j\, ,\ \
R_{+}^{\,i}(kr+j,0)=\left\{\begin{array}{cc}0,& i>j,\\1,&i=j.\end{array}
\right.
\eeq
\end{lem}
Equations (\ref{as4}) are equivalent to the Riemann factorization
problem on $\G$. The dimension of the space of its meromorphic solutions
with poles at $\g_s$ can be found from the Riemann-Roch theorem.
This dimension equals the number of linear equations (\ref{tu}).

If the functions $\Psi_{\pm}=\Psi_{\pm}(t^{\pm},z)$ depend on
some independent variables $t^{\pm}=(t^{\pm}_j)$, then
the corresponding vector-functions $\psi_n$ are functions of the full set
of the variables $\psi_n=\psi_n(t^+,t^-,Q)$.
Let us define the dependence of $\Psi_{\pm}$ on $t^{\pm}$ such, that
the corresponding Baker-Akhiezer functions give solutions of the $2D$ Toda
lattice.

If we restrict ourselves  to the  construction of the $2D$ Toda equations (\ref{toda}), only, then the dependence
on $t_1^+=\xi, t_1^-=\eta$ is defined by ordinary differential equations. Their
coefficients are the functional parameters mentioned above.

Let $a_i(t_1^+), y_i(t_1^-)$ be a set of arbitrary functions. Then
the germ functions $\Psi_{\pm}$ are defined with the help of the equations
\beq\label{e1}
\p_{t_1^{\pm}}
\Psi_{\pm}=M_{\pm}^{0,1}\Psi_{\pm}, \ \ \Psi_{\pm}(0,z)=1,
\eeq
where the matrices $M_{\pm}^{0,1}(t_1^{\pm},z)$ have the form
\beq\label{e2}
M_{+}^{0,1}=\left(\begin{array}{ccccc}
a_0&1&0&\cdots&0\\
0&a_1&1&\cdots&0\\
\cdot&\cdot&\cdot&\cdot&\cdot\\
0&0&0&\cdots&1\\
z^{-1}&0&0&\cdots&a_{r-1}
\end{array}\right)\ ,\ \
M_{-}^{0,1}=\left(\begin{array}{ccccc}
0&0&\cdots&0&b
_0z^{-1}\\
b_1&0&\cdots&0&0\\
0&b_2&\cdots&0&0\\
\cdot&\cdot&\cdot&\cdot&\cdot\\
0&0&\cdots&b_{r-1}&0\\
\end{array}\right),
\eeq
where
\beq\label{e22}
b_i=e^{y_i-y_{i-1}}, \ y_{-1}=y_{r-1}.
\eeq
\begin{theo} The Baker-Akhiezer vector-function $\psi_n$, corresponding to
a generic set of data
$\{\G,P_{\pm}, z_{\pm}, (\g,\a), a_i,\ y_i\}$ satisfies the equations
\beq\label{as6}
\p_{t_1^+}\psi_n=\psi_{n+1}+v_n\psi_n,\ \
\p_{t_1^-}\psi_n=c_n\psi_{n-1},
\eeq
with the coefficients
$$v_n=\p_{t_1^+}\varphi_n,\ \ c_n=e^{\varphi_n-\varphi_{n-1}},$$
where
\beq\label{as7}
\varphi_{kr+i}=y_i(t_1^-)+\ln R_-^{\, i}(kr+i,0,t_1^+,t_1^-).
\eeq
For each function $f\in \A(\G, P^{\pm})$ with the poles of order  $n_{\pm}$
at the punctures $P^{\pm}$ there is a unique
difference operator $L_f$ of the form (\ref{1}) with
coefficients depending on $t_1^{\pm}$ such that
$L_f\psi^i=f\psi^i.$
\end{theo}
The compatibility condition of equations (\ref{as6}) is equivalent to
the  $2D$ Toda lattice equation
(\ref{toda}).
\begin{cor} The functions $\varphi_n$, given by (\ref{as6}), where
the second term is defined by the evaluation at zero of the regular factor
$R_-$ in (\ref{as4}), are solutions of the $2D$ Toda lattice equations.
\end{cor}
Proof of the Theorem is standard and follows from
comparison of analytical properties on $\G$
of the functions defined by the left and the right hand sides
of (\ref{as6}). The proof does not relay essentially on
the specific form (\ref{e2}) of the matrices $M_{\pm}^{0,1}$.
The Theorem remains true if the matrices $M_{\pm}^{0,1}$ are replaced
by matrices of the form
\beq\label{as8}
\widetilde M_{\pm}^{\, 0,1}=M_{\pm}^{\, 0,1}+m_{\pm}(t_1^{\pm}),
\eeq
where entries $m^{ij}_{\pm}$ do not depend on  $z$ and satisfy
the constraints
\beq\label{as9}
m_+^{ij}=0,\ i<j,\ \ m_-^{ij}=0, i\geq j.
\eeq
However, this extension of a set of the germ-functions provides no new
solutions of the $2D$ Toda lattice equations. Indeed, expansion (\ref{as4})
and constraints (\ref{as5}) are invariant under the transformations
\beq\label{as10}
\widetilde \Psi_{\pm}=g_{\pm}\Psi_{\pm},\ \ \widetilde R_{\pm}=R_{\pm}g_{\pm}^{-1},
\eeq
where $g_-(t_1^-)$ is an upper-triangular matrix, and $g_+(t_1^+)$ is a lower-triangular matrix with
$1$ on the diagonal.
Therefore, these transformations do not change the Baker-Akhiezer function.
They correspond to the gauge transformations
\beq\label{as12}
\widetilde M_{\pm}^{\, 0,1}\longmapsto g_{\pm}^{-1}\p_{t_1^{\pm}}g_{\pm}-
g_{\pm}^{-1}\widetilde M_{\pm}^{\, 0,1}g_{\pm}.
\eeq
In each gauge equivalence class there is a unique matrices
satisfying the constraints  $m_{\pm}=0$.

In order to obtain solutions of the full hierarchy of the $2D$ Toda lattice equations
it is enough to define a dependence of the germ functions $\Psi_{\pm}$
on the times $t_p^{\pm}$ with the help of the differential equations
\beq\label{o1}
\p_{t_p^{\pm}}
\Psi_{\pm}=M_{\pm}^{0,\, p}\, \Psi_{\pm}, \ \ \Psi_{\pm}(0,z)=1,
\eeq
where the matrices $M_{\pm}^{0,i}(t^{\pm},z)$ are polynomial
functions of the variable $z^{-1}$.
Compatibility conditions of (\ref{o1}) for each half of the times
$t_p^+$ or $t_p^-$ are gauge equivalent to one of the $r$-reducation of the
KP hierarchy. Because our mein goal is a construction of solutions
of equation (\ref{toda}), we will present an explicit description of the matrices
$M_{\pm}^{0,\, p},\ p>1,$ and detailed analysis of the corresponding
auxiliary soliton system elsewhere.

\subsection{One-puncture case}

In in the previous case of the separated infinities,
the dependence of one-puncture Baker-Alkhiezer function
on the variables $t_p^{\pm}$ is completely  determined
by the dependence on these variables of the germ function $\Psi_0(n,t,z)$.
The later is defined with the help of the equations
\beq\label{o2}
\p_{t_p^{\pm}}\Psi_{0}(n,t,z)=M_{\pm}^{0,\, p}(n,t,z)\, \Psi_{0}(n,t,z)\, ,
\eeq
where the matrices $M_{\pm}^{0,i}(n,t,z)$ are polynomial functions of $z^{-1}$.
Unlike the case of the separated infinities, even for the construction
of solutions to the $2D$ Toda lattice equation
the matrices $M_{\pm}^{0,1}$ should satisfy the compatibility
condition for equations (\ref{o2}) for $p=1$
and the difference equation
\beq\label{o3}
\Psi_0(n+1,t,z)=\X_0(n,t,z)\Psi_0(n,z)\, ,
\eeq
which is equivalent to the definition of $\Psi_0$ as the Wronsky matrix,
corresponding to solutions of (\ref{op}). The matrix
$\X_0=\left(\X_0^{ij}\right),\ \ r_- \leq i,j< r_+-1,$ has the form
\beq\label{o4}
\X_0=
\left(\begin{array}{ccccc}
0& 1&0&\cdots &0\\
0& 0&1&\cdots &0\\
\cdot&\cdot&\cdot&\cdot&\cdot\\
0& 0&0&\cdots &1\\
\chi_{-r_-}^0& \chi_{-r_-+1}^0&\chi_{-r_-+2}^0&\cdots &\chi_{r_+-1}^0
\end{array}\right),
\eeq
where
\beq\label{o5}
\chi_i^0=z^{-1}\delta_{i,\,0}-f_i^{\,0}(n,t)\,.
\eeq
Let $M_{\pm}^{0,1}$ have the form
\beq\label{o6}
M_+^{0,1}=\X_0+A(n,t),\ \ M_-^{0,1}=B(n,t)\X_0^{-1},
\eeq
where $A$ and $B$ are diagonal matrices
\beq\label{o6a}
A=diag\{a(n-r_-,t),\ldots, a(n+r_+-1,t)\},\ \ B=diag\{b(n-r_-,t),\ldots, b(n+r_+-1,t)\}
\eeq
Then, compatibility of equations (\ref{o2}) for $p=1$
and (\ref{o3}) is equivalent to compatibility of the system
$$\p_{t_1^+}\phi_n=\phi_{n+1}+a(n,t)\phi_n,\  \ \
\p_{t_1^-}\phi_n=b(n,t)\phi_{n-1},$$
and equation (\ref{op}).
Therefore, the functions $y_n(t)$ such that $b(n,t)=e^{y_n-y_{n-1}}$,
are the solution of some reduction of the $2D$ Toda lattice equations onto stationary
points of a linear combination of the hierarchy flows, corresponding to the times
$t_{r_-}^-,\ldots, t_{r_+}^+.$

Let us fix a solution $y_n$ of this reduction and denote the corresponding solution
of the linear system (\ref{o2},\, \ref{o3}) by $\Psi_0(n,t_1^+,t_1^-,z)$.
Then the following statement is valid.

\begin{theo} (\cite{kn}) For each smooth genus $g$ algebraic curve $\G$
with fixed local coordinate $z$ in the neighborhood of a puncture
$P_0$, and each generic set of the Tyurin parameters $(\g,\a)$
of degree $lg$ and rank $l$, there is a unique vector function
$\psi_n(t_1^+,t_1^-, Q)$, with coordinates which outside the puncture
$P_0$ have at most simple poles at the points $\g_s$. Their residues at these
points satisfy relations (\ref{tu1a}). In the neighborhood of $P_0$ the row-vector
$\psi_n$ has the form
\beq\label{o8}
\psi_n=\left(\sum_{s=0}^{\infty}\xi_s(n,t_1^+,t_1^-)z^s\right)
\Psi_0(n,t_1^+,t_1^-,z),
\ \xi_0^i=\delta^i_0\,.
\eeq
The functions $\psi_n$ satisfy the equations
\beq\label{o9}
\p_{t_1^+}\psi_n=\psi_{n+1}+\left(\p_{t_1^+}\varphi_n\right)\psi_n,\ \
\p_{t_1^-}\psi_n=\left(e^{\varphi_n-\varphi_{n-1}}\right)\psi_{n-1},
\eeq
where $\varphi_n$ is given by the formula
\beq\label{o10}
\varphi_{n}=y_n(t_1^+,t_1^-)+\ln \, \left(1+\xi_1^{(-1)}(n,t_1^+,t_1^-)\right),
\eeq
in which $\xi_1^{(-1)}$ is the coordinate with the index $i=-1$ of the vector
$\xi_1$ in (\ref{o8}).
\end{theo}
{\bf Example.} In the case of rank $l=2, \ r_{\pm}=1$ the germ function $\Psi_0$
is defined by any solution of the one-dimensional Toda lattice equation
\beq\label{o12}
\ddot y_n=e^{y_n-y_{n-1}}-e^{y_{n+1}-y_{n}},
\eeq
and has the form
\beq\label{o11}
\Psi_0=\Phi(n,t,z)e^{\,x\,z^{-1}}, \ \ x=t_1^{+}+t_1^-,\ \ t=t_1^+-t_1^-,
\eeq
where $\Phi$ is the Wronsky matrix of the auxiliary linear problem
for (\ref{o12}).

\subsection{Deformations of the Tyurin parameters}
The reconstruction problem of the Baker-Akhiezer vector-function
is equivalent to the linear Riemann problem. As it was mentioned above, this problem
in generic case can not be solved
explicitly. At the same time the corresponding solutions of the $2D$ Toda
lattice can be found with the help of the equations for
the deformation of the Tyurin parameters.

Let $\Psi(n,t,Q)$ be the Wronsky matrix whose rows are
the vector Baker-Akhiezer functions $\psi_{n+j}(t,Q)$.
The corresponding deformation of the Tyurin parameters is defined as follows.
In the generic case $\det \Psi(n,t,Q)$ has $gl$ simple zeros $\g_s(n,t)$,
and $\a_s(n,t)$ are defined as the corresponding left null-vector
\beq\label{F2o}
\a_s(n,t)\Psi(n,t,\g_s(n,t))=0\,.
\eeq
Difference equations describing the dynamics of Tyurin parameters with
respect to the variable $n$, were obtained in Section 2.5.
The equations for the continuous deformations follow from the earlier
results of the authors \cite{kn}.

Let us consider the logariphmic derivative of $\Psi$
\beq\label{i1}
\p_{t_{p}^{\pm}}\Psi=M^{p}_{\pm}\Psi.
\eeq
It is a meromorphic function on $\G$ and outside the puncture
has simple poles at the points $\g_s=\g_s(n,t)$.
In the neighborhood of $\g_s$ it has the form
\beq\label{i2}
M={m_s\a_s\over z-z(\g_s)}+\mu_s+O(z-z(\g_s),
\eeq
where $m_s$ is a vector-column. (For brevity we skip
indices $p,\pm$.) The first two coefficients of expansion (\ref{i2})
define the equations of deformation with respect to the variable
$t=t^{p}_{\pm}$
\beq\label{i3}
\p_t \,z(\g_s)=-{\rm Tr} \, (m_s\a_s)=-(\a_s m_s),\ \
\p_t \a_s=-\a_s\mu_s+\kappa_s\a_s.
\eeq
Here $\kappa_s$ is a constant. Equation (\ref{i3}) is a well-defined dynamical system
on the space of Tyurin parameters, which is the symmetric power
$S^{gl}\left(\G\times CP^{l-1}\right)$.

The compatibility condition
\beq\label{kn}
\p_t \X_n=M_{n+1}\X_n-\X_nM_n
\eeq
of the linear problems
\beq\label {kn1}
\Psi_{n+1}=\X_n\Psi_n, \ \ \ \p_t\Psi_n=M_n\Psi_n,
\eeq
is equivalent to a well-defined  system of non-linear
equations on {\it singular parts} of  $\X_n$ и $M_n$
in the neighborhood of the puncture. Here and below
$\Psi_n=\Psi(n,t,Q),\ \X_n=\X(n,t,Q),\
M_n=M(n,t,Q)$.)

\medskip
\noindent{\bf Discrete analog of the Krichever-Novikov equation.}
Let us consider as an instructive example
the nonlinear equations in the case $l=2$ and $g=1$. Recall, that in this case
the coefficients of the linear system defining
the germ function $\Phi$ in (\ref{o11}) have the form
\beq\label{kn1a}
\X_n^0=\left( \begin{array}{rl} \ \ \ \ \ \ 0,& 1\\ -c^0_{n+1},& k-v^0_{n+1}\end{array}\right),
\ \
M^0_n=\left( \begin{array}{rl} -k+2v_n^0,& 2\\ -2c^0_{n+1},& k\end{array}\right),\ \
k=z^{-1}.
\eeq
The Lax equations for this systems are equivalent to the equations of
the one-dimensional Toda lattice.

The leading terms of the Laurent expansion of the "dressed" matrices $\X_n$ (see. (\ref{pr5}))
and $M_n$ have the same form (\ref{kn1a}) but with different functions $c_n, v_n$.
In particular, the matrix  $M_n$  in the neighborhood of $z=0$ has the form
\beq\label{kn2}
M_n=\left( \begin{array}{ll} 2v_n-k,& 2\\ -2c_{n+1},& k\end{array}\right)+
m_{n}k^{-1}+O(k^{-2}),\
k=z^{-1}.
\eeq
Equations (\ref{kn}) imply
\beq\label{kn3}
\dot c_{n+1}=2 c_{n+1}(v_{n+1}-v_n),\ \
\dot v_{n+1}= 2 (c_{n+2}-c_{n+1})+m_n^{22}-m_{n+1}^{22}.
\eeq
Additional terms $m_n^{ij}$ can be expressed through $c_n,v_n$ and the Tyurin parameters
$\g_n^s, a_n^s$. Our goal is to get a closed system of equations
using the dynamics of Tyurin parameters.

For simplicity we consider {\it the symmetric case}, in which
the constant $c$ in (\ref{p17}) equals zero, $c=0$.
From the definition of  $M_n$, it follows that
\beq\label{kn4}
M_n^{21}=-c_{n+1}+\X_n^{21},\ \ M_n^{22}=v_{n+1}+\X_n^{22}\,.
\eeq
Hence,
\beq\label{kn5}
m_n^{22}=\xi_n^{21}=\wp(\g_n).
\eeq
Substitution of this formula in в (\ref{kn3}) gives
\beq\label{kn6}
\dot v_{n+1}= 2 (c_{n+2}-c_{n+1})+\wp(\g_n)-\wp(\g_{n+1}).
\eeq
Equation (\ref{i3}) implies
\beq\label{kn13a}
\dot \g_n=-{\rm res}_{\g_n}M_n=-{a_n^1+a_n^2\over a_n^1-a_n^2}\ .
\eeq
Therefore, $\dot \g_n$ can be identified with the variables $s_n$, defined in (\ref{kn13}),
and equations (\ref{kn15a}, \ref{kn16a}) can be represented in the form
\bea
4c_{n+1}&=&(\dot \g_n^2-1)F(\g_{n+1},\g_n)F(\g_{n-1},\g_n),\label{kn15}\\
2v_{n+1}&=&\dot \g_n F(\g_{n+1},\g_n)-\dot \g_{n+1} F(\g_{n},\g_{n+1})\label{kn16}.
\eea
Let us present two identities, which will be used below.
\bea
\p_u \ln F(u,v)&=&-F(v,u),\label{kn17}\\
\p_v \ln F(u,v)&=&-F(u,v)+2\zeta(2v)-4\zeta(v),\label{kn18}
\eea
where the elliptic function $F(u,v)$ is given by (\ref{kn10}).
The identities can be verified by comparing
of singularities of their right and left hand sides.
Substitution of (\ref{kn15},\ref{kn16}) into the first equation (\ref{kn3}),
gives with the help of (\ref{kn17}, \ref{kn18}) the equation
\beq
\ddot \g_n=(\dot \g_n^2-1)\left(V(\g_{n},\g_{n+1})+V(\g_{n},\g_{n+1})\right)\ ,
\label{kn19}
\eeq
where
\beq
V(u,v)=\zeta(u+v)+\zeta(u-v)-\zeta(2u).\label{kn20}
\eeq
In the same way it is possible to check
that the substitution of (\ref{kn15},\ref{kn16}) into (\ref{kn6}) gives the
same system (\ref{kn19}).

Equations (\ref{kn19}) are Hamiltonian system of equation with
the Hamiltonian
\beq
H=\sum_{n}\ln \left(\sh^{-2}\left(p_n/2\right)\right)+\ln\left(\wp(x_n-x_{n-1})-
\wp(x_n+x_{n-1})\right).
\label{ham}
\eeq
This system was obtained by one of the authors in \cite{elltoda},
as the solution of reconstruction problem of an integrable system
corresponding to a given family of the spectral curves. Inverse problems of this kind
are natural in the framework of the Seiberg-Witten theory in which
families of curves parameterize moduli of nonequivalent physical vacuum states
in supersymmetric gauge models.

In \cite{elltoda} the system (\ref{kn19}) was called elliptic analog
of the Toda lattice. After a suitable change of variables it coincides
with one of systems obtained in \cite{shab} in the framework of solution of the
classification problem of integrable chains. In \cite{elltoda} the system (\ref{kn19})
was identified with the pole system, describing solutions of the $2D$ Toda lattice that are
elliptic in $x$. An appearance of the same system in the theory of rank 2 solutions
of the $2D$ Toda lattice equation came for authors as a complete surprise.

\end{document}